\documentclass[english, aps,prd,twocolumn,twoside,superscriptaddress,floatfix, nofootinbib]{revtex4-1}

\usepackage[utf8]{inputenc}
\usepackage{amsmath}
\usepackage{geometry}
\usepackage{xcolor}
\usepackage{graphicx}
\usepackage{float}
\usepackage{lipsum, babel}
\usepackage{graphicx}
\usepackage{color}
\usepackage{hyperref}
\usepackage{ifthen,amsmath,amssymb, bm}

\newcommand{\beq} {\begin{equation}}
\newcommand{\eeq} {\end{equation}}
\newcommand{\bal} {\begin{aligned}}
\newcommand{\eal} {\end{aligned}}

\begin{document}

\title{Lower bias, lower noise CMB lensing with foreground-hardened estimators}

\author{Noah Sailer}
\email{nsailer@berkeley.edu}
\affiliation{Berkeley Center for Cosmological Physics, Department of Physics,
University of California, Berkeley, CA 94720, USA}
\author{Emmanuel Schaan}
\email{eschaan@lbl.gov}
\affiliation{Lawrence Berkeley National Laboratory, One Cyclotron Road, Berkeley, CA 94720, USA}
\affiliation{Berkeley Center for Cosmological Physics, Department of Physics,
University of California, Berkeley, CA 94720, USA}
\author{Simone Ferraro}
\email{sferraro@lbl.gov}
\affiliation{Lawrence Berkeley National Laboratory, One Cyclotron Road, Berkeley, CA 94720, USA}
\affiliation{Berkeley Center for Cosmological Physics, Department of Physics,
University of California, Berkeley, CA 94720, USA}

\begin{abstract}
Extragalactic foregrounds in temperature maps of the Cosmic Microwave Background (CMB) severely limit the ability of standard estimators to reconstruct the weak lensing potential. 
These foregrounds are not fully removable by multi-frequency cleaning or masking and can lead to large biases if not properly accounted for.  
For foregrounds made of a number of unclustered point sources, an estimator for the source amplitude can be derived and deprojected, removing any bias to the lensing reconstruction.  
We show with simulations that all of the extragalactic foregrounds in temperature can be approximated by a collection of sources with identical profiles, and that a simple bias hardening technique is effective at reducing any bias to lensing, at a minimal noise cost. We compare the performance and bias to other methods such as ``shear-only'' reconstruction, and discuss how to jointly deproject any arbitrary number of foregrounds, each with an arbitrary profile.
In particular, for a Simons Observatory-like experiment foreground-hardened estimators allow us to extend the maximum multipole used in the reconstruction, increasing the overall statistical power by $\sim 50\%$ over the standard quadratic estimator, both in auto and cross-correlation. 
We conclude that source hardening outperforms the standard lensing quadratic estimator both in auto and cross-correlation, and in terms of lensing signal-to-noise and foreground bias.

\end{abstract}

\maketitle

\section{Introduction}
CMB lensing \cite{2006PhR...429....1L, 2010GReGr..42.2197H} is rapidly becoming one of the most powerful tools in a cosmologist's toolkit. With the ability to measure matter fluctuations to high redshift, no photometric redshift uncertainties and its sensitivity to crucial cosmological parameters, sub-percent precision measurements have the potential to greatly increase our understanding of the Universe.

Current and upcoming wide-field CMB experiments such as AdvACT \cite{2016JLTP..184..772H}, SPT-3G \cite{2014SPIE.9153E..1PB} and Simons Observatory \cite{2019JCAP...02..056A} will heavily rely on reconstruction from CMB temperature fluctuations (as opposed to polarization). 
For future polarization-dominated experiments like CMB-S4 \cite{2019arXiv190704473A} and CMB-HD \cite{sehgal2019cmbhd,sehgal2020cmbhd}, temperature lensing will still contribute a non-negligible part of the total signal-to-noise.

Temperature maps contain significant contamination from extragalactic foregrounds such as the thermal and kinematic Sunyaev-Zel'dovich effects (tSZ and kSZ), the Cosmic Infrared Background (CIB), and both radio and infrared point sources. Lensing reconstruction on contaminated maps creates significant biases in both the auto- and cross-correlation of the reconstructed field, which in turn can bias our inference of cosmology \cite{2014ApJ...786...13V,2014JCAP...03..024O, 2018PhRvD..98b3534M,2018PhRvD..97b3512F,2019PhRvL.122r1301S,2019PhRvD..99b3508B}.
These biases can be up to several tens of percent for upcoming surveys, thus severely limiting our ability to use CMB lensing for precision cosmology.

Several techniques have been proposed and implemented to mitigate the impact of extragalactic foregrounds.
Some are based on symmetries, such as the recently proposed shear-only reconstruction \cite{2019PhRvL.122r1301S} (and related generalizations such as hybrid and general multipole estimators), while others make use of multi-frequency maps to reduce the impact of particular foregrounds \cite{2018PhRvD..98b3534M, 2020arXiv200401139D}.

Here we revisit and extend a bias hardening technique first proposed in \cite{2014JCAP...03..024O, 2013MNRAS.431..609N} and applied to real data in \cite{plancklensing2013}, and compare it to shear-only reconstruction and the hybrid estimator of \cite{2019PhRvL.122r1301S}. We show that if we approximate extragalactic foregrounds as a collection of sources with identical profiles, an estimator for their amplitude can be obtained and hardened against in an optimal way to mitigate their impact on the reconstructed lensing convergence. With the use of realistic and correlated simulations of all of the foregrounds, we show that this approximation is excellent and leads to a very large suppression in bias, allowing for the use of considerably smaller scales in the reconstruction.

The remainder of the paper is organized as follows:  In Section \ref{sec:qe_bh} we review the general formalism for bias hardening against a single foreground, then generalize the procedure to an arbitrary number of foregrounds, each with arbitrary profiles. In Section \ref{sec:qe_source} we specialize to the case where foregrounds are point sources, and in Section \ref{sec:biases} we numerically explore the biases and the performance of mitigation through bias hardening or shear-only estimators.  We conclude with Section \ref{sec:conclusions}.                                    
\section{Foreground-hardened quadratic estimators}
\label{sec:qe_bh}
In this section, we first review the derivation of the standard lensing quadratic estimator (QE),
and rephrase it in terms of the linear response of the map covariance to the lensing convergence.
We build on this formalism to construct foreground estimators, and finally null the linear response of the lensing QE to these foregrounds, as in \cite{2014JCAP...03..024O, 2013MNRAS.431..609N}.

\subsection{Lensing quadratic estimator, bispectrum \& linear response}
The observed CMB temperature $T$ is the sum of the \textit{lensed} primary CMB $T^\text{CMB}$ and the foregrounds $s$, which will be approximated by a collection of sources in this work: $T = T^\text{CMB} + s$.
The lensed CMB temperature $T^\text{CMB}$ field is statistically invariant under translations.
As a result, the off-diagonal covariance $\langle T^\text{CMB}_{\boldsymbol{\ell}} T^\text{CMB}_{\boldsymbol{L}-\boldsymbol{\ell}} \rangle$ with $\boldsymbol{L}\neq 0$ is zero.
However, fixing the lensing convergence Fourier mode $\kappa_{\bm{L}}$ breaks this statistical homogeneity, and produces non-zero off-diagonal covariances:
\beq
\bal
\langle T^\text{CMB}_{\boldsymbol{\ell}} T^\text{CMB}_{\boldsymbol{L}-\boldsymbol{\ell}} \rangle_{\text{at fixed }\kappa_{\boldsymbol{L}}}
=
f^\kappa_{\boldsymbol{\ell},\boldsymbol{L}-\boldsymbol{\ell}}\kappa_{\boldsymbol{L}}
+
\mathcal{O}\left( \kappa_{\boldsymbol{L}}^2 \right)\\
\quad\text{with}\quad
f^\kappa_{\boldsymbol{\ell},\boldsymbol{L}-\boldsymbol{\ell}} 
\equiv
\frac{2\boldsymbol{L}}{L^2}
\cdot\left[\boldsymbol{\ell}C^0_\ell+(\boldsymbol{L}-\boldsymbol{\ell})C^0_{|\boldsymbol{L}-\boldsymbol{\ell}|}
\right]
,
\label{eq:lensing_response}
\eal
\eeq
where $C^0_\ell$ is the \textit{unlensed} power spectrum. From this, we obtain an unbiased lensing estimator (to first order in $\kappa$) 
from the ratio
$T_{\boldsymbol{\ell}} T_{\boldsymbol{L}-\boldsymbol{\ell}} / f^\kappa_{\boldsymbol{\ell},\boldsymbol{L}-\boldsymbol{\ell}} $.
The standard lensing QE \cite{2002ApJ...574..566H} is simply the inverse-variance weighted average of all these ``building block'' estimators, summing over $\boldsymbol{\ell}$ at fixed $\boldsymbol{L}$. This leads to the usual result:
\beq
\bal
    \hat{\kappa}_{\boldsymbol{L}} &= N^\kappa_{\boldsymbol{L}}\int \frac{d^2\boldsymbol{\ell}}{(2\pi)^2} F^\kappa_{\boldsymbol{\ell},\boldsymbol{L}-\boldsymbol{\ell}} T_{\boldsymbol{\ell}} T_{\boldsymbol{L}-\boldsymbol{\ell}}
    \\
    \left(N^\kappa_{\boldsymbol{L}}\right)^{-1} &= \int \frac{d^2\boldsymbol{\ell}}{(2\pi)^2} F^\kappa_{\boldsymbol{\ell},\boldsymbol{L}-\boldsymbol{\ell}} f^\kappa_{\boldsymbol{\ell},\boldsymbol{L}-\boldsymbol{\ell}}
    \\
    F^\kappa_{\boldsymbol{\ell},\boldsymbol{L}-\boldsymbol{\ell}} &= \frac{f^\kappa_{\boldsymbol{\ell},\boldsymbol{L}-\boldsymbol{\ell}}}{2 C^\text{tot}_{\ell}C^\text{tot}_{|\boldsymbol{L}-\boldsymbol{\ell}|}}.
\label{eq:standard_qe_kappa}
\eal
\eeq
In order to generalize this construction to foreground quadratic estimators,
we consider Eq.~\eqref{eq:lensing_response} and note that 
$f^\kappa_{\boldsymbol{\ell},\boldsymbol{L}-\boldsymbol{\ell}}$
can be seen as the linear response of the two point correlator 
$
\langle T_{\bm{\ell}} T_{\boldsymbol{L}-\boldsymbol{\ell}} \rangle_{\text{at fixed }\kappa_{\boldsymbol{L}} }
$
to the lensing convergence mode $\kappa_{\boldsymbol{L}}$:
\beq
f^\kappa_{\boldsymbol{\ell},\boldsymbol{L}-\boldsymbol{\ell}}
=
\frac{\delta \langle T^\text{CMB}_{\boldsymbol{\ell}} T^\text{CMB}_{\boldsymbol{L}-\boldsymbol{\ell}} \rangle_{\text{at fixed }\kappa_{\boldsymbol{L}}}}
{\delta \kappa_{\boldsymbol{L}}}
(\kappa_{\boldsymbol{L}}=0).
\eeq
Furthermore, by multiplying both sides of Eq.~\eqref{eq:lensing_response} with $\kappa_{-\boldsymbol{L}}$, we see that this linear response can be expressed as the ratio of the following bispectrum and power spectrum:
\beq
\bal
f^\kappa_{\boldsymbol{\ell},\boldsymbol{L}-\boldsymbol{\ell}}
=
\frac{\langle T^\text{CMB}_{\boldsymbol{\ell}} T^\text{CMB}_{\boldsymbol{L}-\boldsymbol{\ell}} \kappa_{-{\boldsymbol{L}}} \rangle}
{\langle \kappa_{\boldsymbol{L}} \kappa_{-\boldsymbol{L}} \rangle}.
\eal
\eeq
This final expression is our starting point to derive foreground quadratic estimators.

\subsection{Foreground quadratic estimators}

Following the argument above, anytime a field $s$ (e.g. a foreground, the mask, the beam) has a non-zero bispectrum $\langle s_{\boldsymbol{\ell}} s_{\boldsymbol{L}-\boldsymbol{\ell}} s_{-\boldsymbol{L}} \rangle$,
we can define the following linear response\footnote{We note that this expression only accounts for the non-Gaussianity of the sources. In principle, the sources are themselves lensed which results in an additional smaller correlation and $\kappa$ dependence of the mask \cite{2019PhRvD.100l3504M}. For simplicity, we ignore this effect in this work.}:
\beq
\bal
f^s_{\boldsymbol{\ell},\boldsymbol{L}-\boldsymbol{\ell}}
&= \frac{\langle s_{\boldsymbol{L}} s_{\boldsymbol{L}-\boldsymbol{\ell}} s_{-{\boldsymbol{L}}} \rangle}
{\langle s_{\boldsymbol{L}} s_{-\boldsymbol{L}} \rangle}
\eal
\eeq
such that
\beq
\langle s_{\boldsymbol{\ell}} s_{\boldsymbol{L}-\boldsymbol{\ell}} \rangle_{\text{at fixed }s_{\boldsymbol{L}}}
=
f^s_{\boldsymbol{\ell},\boldsymbol{L}-\boldsymbol{\ell}}
s_{\boldsymbol{L}}
+
\mathcal{O}\left( (s_{\boldsymbol{L}})^2 \right).
\eeq
We can then define the ``building block'' foreground estimator, unbiased to first order in the foreground $s$, via 
$T_{\boldsymbol{\ell}} T_{\boldsymbol{L}-\boldsymbol{\ell}} / f^s_{\boldsymbol{\ell},\boldsymbol{L}-\boldsymbol{\ell}}$.
Finally, we inverse-variance weight these estimators for all $\boldsymbol{\ell}$, at fixed $\boldsymbol{L}$, to obtain the standard minimum variance quadratic estimator for $s$. 
In practice, evaluating this estimator requires knowledge of the foreground non-Gaussianity, through its bispectrum.
We note however that no assumption about the foreground trispectrum is used.

\subsection{Bias-hardened lensing quadratic estimators}
In this section we review the derivation of the bias-hardened (BH) lensing estimator of \cite{2014JCAP...03..024O, 2013MNRAS.431..609N} for a single foreground, relate the noise of the BH estimator to the noise of the standard minimum variance QE, and generalize the procedure for an arbitrary number of foregrounds.

\subsubsection{Hardening against a single foreground}
In the presence of a single foreground $s$, the off-diagonal covariances of the observed temperature map take the form 
\beq
\langle T_{\boldsymbol{\ell}} T_{\boldsymbol{L}-\boldsymbol{\ell}} \rangle
=
f^\kappa_{\boldsymbol{\ell},\boldsymbol{L}-\boldsymbol{\ell}}\kappa_{\boldsymbol{L}}
+
f^s_{\boldsymbol{\ell},\boldsymbol{L}-\boldsymbol{\ell}}s_{\boldsymbol{L}}
\eeq
to lowest order in $\kappa$ and $s$. In this equation the average is taken at fixed $\kappa_{\boldsymbol{L}}$ and $s_{\boldsymbol{L}}$, which will be implied from here on out for notational economy. Since both $\kappa_{\boldsymbol{L}}$ and $s_{\boldsymbol{L}}$ produce off-diagonal covariances, the standard quadratic estimators for $\kappa$ and $s$ acquire a bias. These biases can be written in the form \cite{2014JCAP...03..024O}:
\beq
    \begin{pmatrix}
     \langle\hat{\kappa}_{\boldsymbol{L}}\rangle\\
     \langle\hat{s}_{\boldsymbol{L}}\rangle
    \end{pmatrix} =
    \begin{pmatrix}
    1 & N^\kappa_{\boldsymbol{L}}\mathcal{R}_{\boldsymbol{L}}\\
    N^{s}_{\boldsymbol{L}}\mathcal{R}_{\boldsymbol{L}} & 1
    \end{pmatrix}
    \begin{pmatrix}
     \kappa_{\boldsymbol{L}}\\
     s_{\boldsymbol{L}}
    \end{pmatrix}
\label{eq:responses_k_s}
\eeq
where the response $\mathcal{R}_{\boldsymbol{L}}$ is defined to be
\beq
\bal
    \mathcal{R}_{\boldsymbol{L}}  &= \int \frac{d^2\boldsymbol{\ell}}{(2\pi)^2}
    \frac{
    f^\kappa_{\boldsymbol{\ell},\boldsymbol{L}-\boldsymbol{\ell}} 
    f^s_{\boldsymbol{\ell},\boldsymbol{L}-\boldsymbol{\ell}} 
    }{
    2 C^\text{tot}_\ell C^\text{tot}_{|\boldsymbol{L}-\boldsymbol{\ell}|}
    }
    .
\eal
\eeq
We see that the bias to $\hat{\kappa}$ is proportional to $s$, and vice-versa. This allows us to create new bias-hardened estimators, which are the unique linear combinations of $\hat{\kappa}$ and $\hat{s}$ that null these biases 
\begin{equation}
    \begin{pmatrix}
     \hat{\kappa}^\text{BH}_{\boldsymbol{L}}\\
     \hat{s}^{\text{BH}}_{\boldsymbol{L}}
    \end{pmatrix} =
    \begin{pmatrix}
    1 & N^\kappa_{\boldsymbol{L}}\mathcal{R}_{\boldsymbol{L}}\\
    N^{s}_{\boldsymbol{L}}\mathcal{R}_{\boldsymbol{L}} & 1
    \end{pmatrix}^{-1}
    \begin{pmatrix}
     \hat{\kappa}_{\boldsymbol{L}}\\
     \hat{s}_{\boldsymbol{L}}
    \end{pmatrix}
    .
\label{eq:bias-hardened_k_s}
\end{equation}
By taking an average of Eq.~\eqref{eq:bias-hardened_k_s}, and inserting Eq.~\eqref{eq:responses_k_s} on the right hand side, we see that bias-hardened estimators indeed satisfy 
$\langle \hat{\kappa}^\text{BH}_{\boldsymbol{L}}\rangle = \kappa_{\boldsymbol{L}}$ 
and 
$\langle \hat{s}^\text{BH}_{\boldsymbol{L}}\rangle = s_{\boldsymbol{L}}$.

\subsubsection{Noise of the hardened lensing QE}

By construction, the weights of the standard QE are chosen to minimize its variance. Since the BH estimator's weights differ from the standard QE's, bias-hardening comes at a price in noise, which we quantify in this section. We start with Eq.~\eqref{eq:bias-hardened_k_s}, the definition of the BH estimator, from which we compute the variance
\beq
\bal
    \left\langle \hat{\kappa}^\text{BH}_{\boldsymbol{L}} \hat{\kappa}^\text{BH}_{-\boldsymbol{L}} \right\rangle 
    &= 
    \bigg[
    \left\langle \hat{\kappa}_{\boldsymbol{L}} \hat{\kappa}_{-\boldsymbol{L}} \right\rangle  +
    \left(N^{\kappa}_{\boldsymbol{L}}\mathcal{R}_{\boldsymbol{L}} \right)^2 
    \left\langle \hat{s}_{\boldsymbol{L}} \hat{s}_{-\boldsymbol{L}} \right\rangle 
    \\
    &-
    2N^{\kappa}_{\boldsymbol{L}}\mathcal{R}_{\boldsymbol{L}} \text{Re}
    \left\langle
    \hat{\kappa}_{\boldsymbol{L}}\hat{s}_{-\boldsymbol{L}}
    \right\rangle
    \bigg]
    \bigg/
    \left(1- N^{\kappa}_{\boldsymbol{L}}N^{s}_{\boldsymbol{L}}\mathcal{R}^2_{\boldsymbol{L}} \right)^{-2}.
\label{eq:bh_noise_derivation}
\eal
\eeq
In this equation we've used the identity $N^{\kappa}_{\boldsymbol{L}}=N^{\kappa}_{-\boldsymbol{L}}$, and we've assumed that $f^s_{\bm{\ell},\bm{L}-\bm{\ell}}$ is even in both of its arguments, which forces both the noise of the source estimator and the response to be even functions of $\bm{L}$. This assumption is valid when modeling the field $s$ as a collection of individual sources with
profiles $u_{\bm{\ell}}$ (since $u_{\bm{\ell}} = u_{-\bm{\ell}}$ for any physical emission profile), as in section \ref{sec:qe_source}.

\begin{figure}[!h]
\centering
\includegraphics[width=\linewidth]{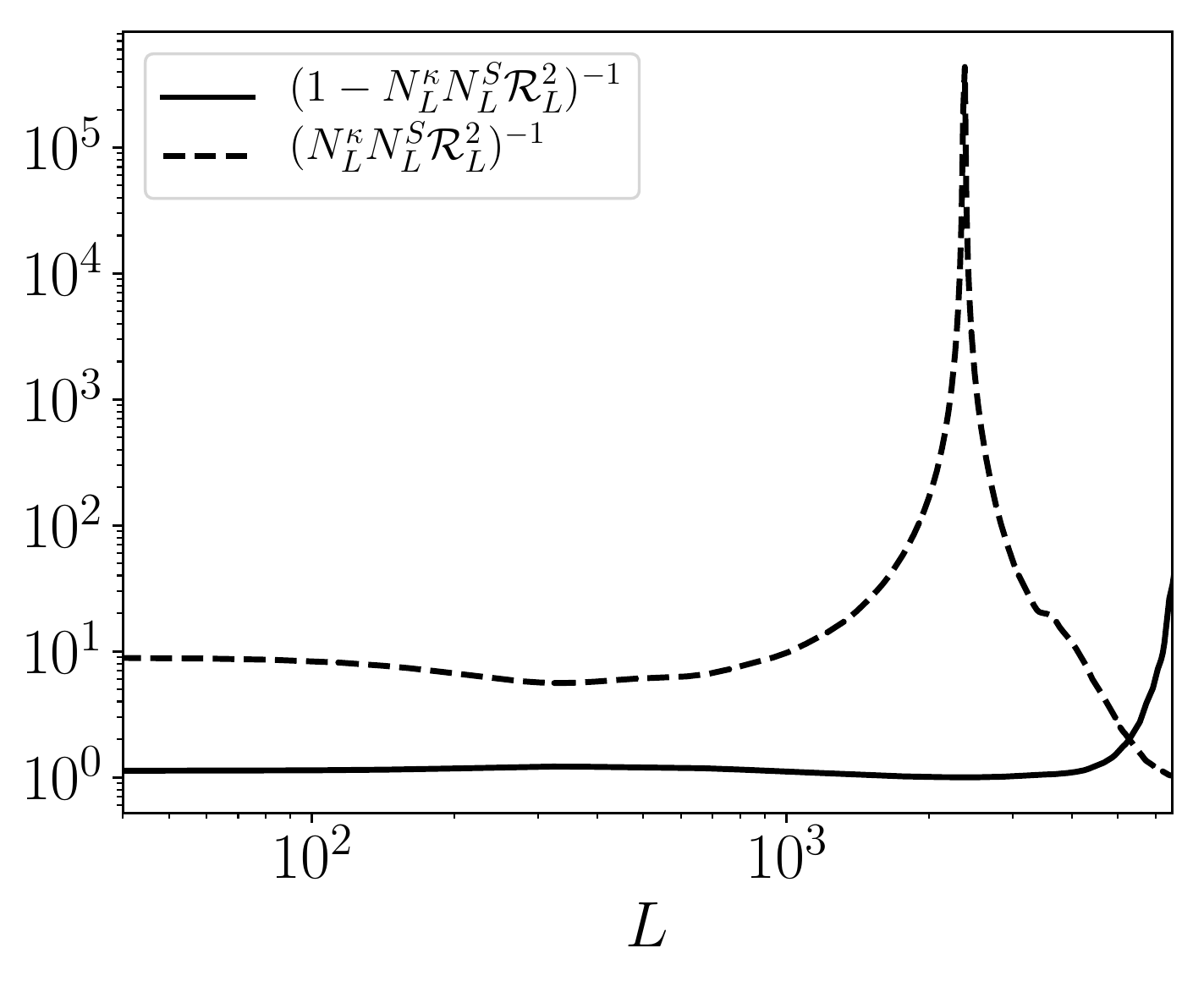}
\caption{
Bias-hardening against point source foregrounds. 
The solid curve is the determinant of the matrix in Eq. \eqref{eq:bias-hardened_k_s}, which is the ratio of the bias-hardened estimator's noise to the standard QE's noise. 
It is close to unity on most scales, implying that the noise penalty for bias hardening is minimal.
The dashed curve is the ratio of the SNR of the point source estimator to the relative bias to $C^\kappa_L$ from the source trispectrum. This ratio is roughly $10$ at all scales. Thus in the regime where the bias to the standard QE is significant, the source estimator reconstructs the sources with a high SNR, and the source-hardened estimator can remove the bias due to sources with little noise cost. The spike is due to a zero crossing in the response $\mathcal{R}_L$ at $L\sim 2500$.
}
\label{fig:determinant}
\end{figure}

The cross-correlation of $\hat{\kappa}$ and $\hat{s}$ is proportional to the response function, and is explicitly given by
\beq
    \big\langle
    \hat{\kappa}_{\boldsymbol{L}}\hat{s}_{\boldsymbol{L}'}
    \big\rangle 
    = 
    (2\pi)^2
    \delta^D_{\boldsymbol{L}+\boldsymbol{L}'}N^\kappa_{\boldsymbol{L}}N^{s}_{\boldsymbol{L}}\mathcal{R}_{\boldsymbol{L}}
    .
\label{eq:cross_k_s}
\eeq
Combining Eq.~\eqref{eq:bh_noise_derivation} and~\eqref{eq:cross_k_s} results in the simple expression
\beq
    N^{\kappa^\text{BH}}_{\boldsymbol{L}}
    = 
    N^\kappa_{\boldsymbol{L}}\left(1- N^{\kappa}_{\boldsymbol{L}}N^{s}_{\boldsymbol{L}}\mathcal{R}^2_{\boldsymbol{L}} \right)^{-1}.
\eeq
We see that the noise cost for bias-hardening is completely determined by the determinant of the matrix in Eq.~\eqref{eq:bias-hardened_k_s}.
The value of this determinant is plotted in Fig.~\ref{fig:determinant} for the case of point source foregrounds ($f^s_{\bm{\ell},\bm{L}-\bm{\ell}}=1$).
Nearly identical calculations yield the noise of the bias-hardened source estimator and the noise cross-correlation of the two bias-hardened estimators. The results are
\beq
\bal
    N^{s^\text{BH}}_{\boldsymbol{L}} &= N^{s}_{\boldsymbol{L}} \left(1- N^{\kappa}_{\boldsymbol{L}}N^{s}_{\boldsymbol{L}}\mathcal{R}^2_{\boldsymbol{L}} \right)^{-1}\\
    \langle\hat{\kappa}^\text{BH}_{\boldsymbol{L}} \hat{s}^\text{BH}_{\boldsymbol{L}}\rangle &=
    -
    \langle\hat{\kappa}_{\boldsymbol{L}} \hat{s}_{\boldsymbol{L}}\rangle
    \left(1-N^{\kappa}_{\boldsymbol{L}}N^{s}_{\boldsymbol{L}}\mathcal{R}^2_{\boldsymbol{L}}\right)^{-1}
    .
\eal
\eeq

\subsubsection{Simultaneously hardening against $n$ foregrounds}

It is straightforward to generalize the bias-hardening procedure for $n$ foreground fields $s_{1,\boldsymbol{L}},\cdots, s_{n,\boldsymbol{L}}$. In this case the observed map has off-diagonal covariances
\beq
    \langle T_{\boldsymbol{\ell}} T_{\boldsymbol{L}-\boldsymbol{\ell}}\rangle
    =
    f^\kappa_{\boldsymbol{\ell},\boldsymbol{L}-\boldsymbol{\ell}}
    \kappa_{\boldsymbol{L}}
    +
    f^{s_1}_{\boldsymbol{\ell},\boldsymbol{L}-\boldsymbol{\ell}}
    s_{1,\boldsymbol{L}}
    +
    \cdots
    +
    f^{s_n}_{\boldsymbol{\ell},\boldsymbol{L}-\boldsymbol{\ell}}
    s_{n,\boldsymbol{L}}
\eeq
to lowest order in $\kappa$ and the foreground fields. Just as before, this average is taken at fixed $\kappa_{\bm{L}},s_{1,\bm{L}},\cdots,s_{n,\bm{L}}$. First, we write down the standard minimum variance quadratic estimators for each of the fields: 
\beq
\bal
    \hat{X}_{\boldsymbol{L}} &= N^X_{\boldsymbol{L}}\int \frac{d^2\boldsymbol{\ell}}{(2\pi)^2} F^X_{\boldsymbol{\ell},\boldsymbol{L}-\boldsymbol{\ell}} T_{\boldsymbol{\ell}} T_{\boldsymbol{L}-\boldsymbol{\ell}}
    \\
    \left(N^X_{\boldsymbol{L}}\right)^{-1} &= \int \frac{d^2\boldsymbol{\ell}}{(2\pi)^2} F^X_{\boldsymbol{\ell},\boldsymbol{L}-\boldsymbol{\ell}} f^X_{\boldsymbol{\ell},\boldsymbol{L}-\boldsymbol{\ell}}
    \\
    F^X_{\boldsymbol{\ell},\boldsymbol{L}-\boldsymbol{\ell}} &= \frac{f^X_{\boldsymbol{\ell},\boldsymbol{L}-\boldsymbol{\ell}}}{2 C^\text{tot}_{\ell}C^\text{tot}_{|\boldsymbol{L}-\boldsymbol{\ell}|}},
\eal
\eeq
where $X = \kappa, s_1, \cdots, s_n$. We then write down the $(n+1)\times(n+1)$ matrix that quantifies the biases to all the estimators, analogous to Eq.~\eqref{eq:responses_k_s}. Finally, we invert this matrix to obtain the bias-hardened estimators 
\begin{equation}
\footnotesize{
    \begin{pmatrix}
     \hat{\kappa}^\text{BH}_{\boldsymbol{L}}\\
     \hat{s}^\text{BH}_{1,\boldsymbol{L}} \\
     \vdots\\
     \hat{s}^\text{BH}_{n,\boldsymbol{L}}
    \end{pmatrix} =
    \underbrace{
    \begin{pmatrix}
    1 & 
    N^\kappa_{\boldsymbol{L}}
    \mathcal{R}^{\kappa,s_1}_{\boldsymbol{L}} &
    \cdots &
    N^\kappa_{\boldsymbol{L}}
    \mathcal{R}^{\kappa,s_n}_{\boldsymbol{L}}\\
    N^{s_1}_{\boldsymbol{L}}
    \mathcal{R}^{s_1,\kappa}_{\boldsymbol{L}} & 
    1 &
    \cdots
    &
    N^{s_1}_{\boldsymbol{L}}
    \mathcal{R}^{s_1,s_n}_{\boldsymbol{L}}
    \\
    \vdots &
    \vdots
    &
    \vdots
    &
    \vdots\\
    N^{s_n}_{\boldsymbol{L}}
    \mathcal{R}^{s_n,\kappa}_{\boldsymbol{L}}
    &
    N^{s_n}_{\boldsymbol{L}}
    \mathcal{R}^{s_n,s_1}_{\boldsymbol{L}}
    &
    \cdots
    &
    1
    \end{pmatrix}^{-1}}_{\equiv \bm{M}^{-1}_{\bm{L}}}
    \begin{pmatrix}
     \hat{\kappa}_{\boldsymbol{L}}\\
     \hat{s}_{1,\boldsymbol{L}} \\
     \vdots\\
     \hat{s}_{n,\boldsymbol{L}}
    \end{pmatrix}
}
\label{eq:generalizedBHMatrix}
\end{equation}
where the generalized response is defined to be 
\begin{equation}
    \mathcal{R}^{X,Y}_{\boldsymbol{L}} 
    =
    \int \frac{d^2\boldsymbol{\ell}}{(2\pi)^2}
    \frac{
    f^X_{\boldsymbol{\ell},\boldsymbol{L}-\boldsymbol{\ell}}
    f^Y_{\boldsymbol{\ell},\boldsymbol{L}-\boldsymbol{\ell}}
    }{
    2 C^\text{tot}_\ell C^\text{tot}_{|\boldsymbol{L}-\boldsymbol{\ell}|}}
    .
\end{equation}
Just as in the previous section, the response $\mathcal{R}^{X,Y}_{\boldsymbol{L}}$ is proportional to the cross-correlation of the two fields $X$ and $Y$. When $X=Y$, the response is simply the inverse of the noise, which is manifested in Eq. \eqref{eq:generalizedBHMatrix} by the 1's on the diagonal. Writing the matrix $\bm{M}_{\bm{L}}$ as 
\beq
\bm{M}_{\bm{L}}=
\begin{pmatrix}
1 & \bm{w}^T_{\bm{L}} \\
\bm{v}_{\bm{L}} & \bm{A}_{\bm{L}}
\end{pmatrix},
\eeq
where $\bm{A}_{\bm{L}}$ is a $n\times n$ matrix and $\bm{w}_{\bm{L}},\bm{v}_{\bm{L}}$ are $n$-dimensional column vectors, we find that the noise of the BH lensing estimator is simply
\beq
N^{\kappa^\text{BH}}_{\bm{L}}
=
\frac{\det(\bm{A}_{\bm{L}})}{\det(\bm{M}_{\bm{L}})}
N^{\kappa}_{\bm{L}}.
\label{eq:general_noise}
\eeq
A derivation of this equation is given in Appendix~\ref{app:derivation of general noise}.
\section{Source QE and source-hardened lensing QE}
\label{sec:qe_source}

\subsection{Halo model for sources}

In the previous section we reviewed how to bias-harden against any number of foregrounds. The only input to this procedure is the linear response $f^s_{\bm{\ell},\bm{L}-\bm{\ell}}$ of each foreground, which can be calculated if the foreground's bispectrum and power spectrum are known. To estimate these quantities we adopt a halo model, writing each foreground $s(\bm{x})$ as a sum of individual sources with profiles $u(\bm{x})$:
\beq
s(\boldsymbol{x}) = \sum_i s_i u(\boldsymbol{x} - \boldsymbol{x}_i),
\eeq
where $s_i$ and $\boldsymbol{x}_i$ are the flux and position of the $i$'th source, and the profile $u(\boldsymbol{x})$ is assumed to be source independent.
These sources are assumed to be a Poisson sampling of the matter density field.
Because they trace matter on large scales, the source field $s$ is correlated with the true lensing convergence field. 
As we show below, this correlation causes a bias in the CMB lensing cross- and auto-correlations.
With this model, the linear response takes the form
\beq
f^{s}_{\boldsymbol{\ell},\boldsymbol{L}-\boldsymbol{\ell}}
=
\frac{\langle s_{\boldsymbol{\ell}} s_{\boldsymbol{L}-\boldsymbol{\ell}} s_{-\boldsymbol{L}} \rangle}
{\langle s_{\boldsymbol{L}} s_{-\boldsymbol{L}} \rangle}
=
\frac{\langle s^3_i \rangle}{\langle s^2_i\rangle}
\frac{ u_{\boldsymbol{\ell}}  
u_{\boldsymbol{L}-\boldsymbol{\ell}}}
{  u_{\boldsymbol{L}}},
\label{eq:linear response to sources}
\eeq
where in the second equality we assumed that the sources are sparse, 
such that the bispectrum and power spectrum are both shot noise dominated, and the source clustering is negligible.
In this case the linear response simplifies to a separable function of $\boldsymbol{\ell}$ and $\boldsymbol{L}-\boldsymbol{\ell}$, and is proportional to the ratio of the third to the second moment of the individual source fluxes. 

We can choose the normalization of the source estimator $\hat{s}$ to remove the constant factor $\langle s^3_i \rangle / \langle s^2_i \rangle$. Doing so makes $\hat{s}$ an estimator of $s'\equiv s\langle s^2_i \rangle / \langle s^3_i \rangle$, which has linear response 
$
f^{s'}_{\bm{\ell},\bm{L}-\bm{\ell}}
=
u_{\bm{\ell}}
u_{\bm{L}-\bm{\ell}}
/
u_{\bm{L}}
$.
This simplifies the expression of the source estimator to
\beq
\bal
    \hat{s}_{\boldsymbol{L}} &=  N^{s}_{\boldsymbol{L}}\int \frac{d^2\boldsymbol{\ell}}{(2\pi)^2} \frac{T_{\boldsymbol{\ell}} T_{\boldsymbol{L}-\boldsymbol{\ell}}}{2 C^\text{tot}_{\ell}C^\text{tot}_{|\boldsymbol{L}-\boldsymbol{\ell}|}}
    \frac{u_{\boldsymbol{\ell}}
    u_{\boldsymbol{L}-\boldsymbol{\ell}}}
    {u_{\boldsymbol{L}}}
    \\
    \left(N^{s}_{\boldsymbol{L}}\right)^{-1} &= \int \frac{d^2\boldsymbol{\ell}}{(2\pi)^2} \frac{1}{2 C^\text{tot}_{\ell}C^\text{tot}_{|\boldsymbol{L}-\boldsymbol{\ell}|}}
    \left[
    \frac{u_{\boldsymbol{\ell}}
    u_{\boldsymbol{L}-\boldsymbol{\ell}}}
    {u_{\boldsymbol{L}}}
    \right]^2
    ,
\eal
\eeq
which can be quickly evaluated using FFT. We emphasize that the weights and normalization of the BH estimator are left unchanged by this renormalization of $\hat{s}$. Therefore the bias-hardening procedure requires no knowledge of the source amplitude, and the normalization of the profile $u$ is completely arbitrary. For this reason we will drop the primes from here-on-out, defining 
$
f^{s}_{\bm{\ell},\bm{L}-\bm{\ell}}
=
u_{\bm{\ell}}
u_{\bm{L}-\bm{\ell}}
/
u_{\bm{L}}
$.

\subsection{Noise cost and importance of the non-Gaussianity}

Consider a single source component $s$.
If $s$ were Gaussian, it would be indistinguishable from map noise, and the source estimator $\hat{s}$ would not be able to pick it out.
Here, we show how the non-Gaussianity of the source map controls both the signal-to-noise (SNR) of the source estimator, and the bias from sources to CMB lensing.
For simplicity let $s$ be a Poisson distributed point source field ($f^s_{\bm{\ell},\bm{L}-\bm{\ell}}=u_{\boldsymbol{\ell}}=1$). In this case the mean number of sources $\bar{n}$ determines the non-Gaussianity; as $\bar{n}$ goes to infinity, the statistics converges to that of a Gaussian.

The point source estimator is an unbiased (to lowest order) estimator of $s$, so the SNR per Fourier mode is simply :
\begin{align}
    \text{SNR}^s_L &= \langle \hat{s}_{\boldsymbol{L}} \hat{s}_{-\boldsymbol{L}}\rangle / N^s_L\\
                   &= \mathcal{T}^{s} /N^s_L.
\label{eq:SNR_from_trispectrum}
\end{align}
Thus the SNR for the source estimator scales with the scale-independent source trispectrum $\mathcal{T}^s$: it is larger for low source densities which corresponds to a higher non-Gaussianity.
As the number of sources increases, the field becomes increasingly Gaussian, and can no longer be distinguished from Gaussian noise in the map.
The expression above is valid for mildly non-Gaussian source field, as is the case in reality. For the very highly non-Gaussian case, which is not relevant for the CMB, the source non-Gaussianity contributes also to the noise, leading to a saturation of the SNR to a finite value as $\bar{n}$ goes to zero.

Even if the source field and lensing convergence were independent, point sources would cause a bias to the standard QE due to its trispectrum:
\begin{equation}
\begin{aligned}
    \frac{\text{bias to }C_L^{\kappa}\text{ from }\mathcal{T}^s}{N^\kappa_L}
    &= 
    \frac{\mathcal{T}^{s}}{N^\kappa_L}
    \left(
    N^\kappa_L
    \int \frac{d^2\bm{\ell}}{(2\pi)^2}
    F^\kappa_{\bm{\ell},\bm{L}-\bm{\ell}}
    \right)^2
    \\
    &=
    \mathcal{T}^{s}N^\kappa_L \mathcal{R}^2_L.
\label{eq:bias_to_QE_from_trispectrum}
\end{aligned}
\end{equation}
Here again, the source non-Gaussianity controls the bias to lensing via the $1/\bar{n}$ scaling. From equations \eqref{eq:SNR_from_trispectrum} and \eqref{eq:bias_to_QE_from_trispectrum}, we see that the ratio of the SNR of the source estimator to the relative bias to lensing from the source trispectrum is $(N^\kappa_L N^S_L \mathcal{R}^2_L)^{-1}$.
This ratio cancels out the non-Gaussianity of the sources
and is around 10 for most scales of interest, as shown in Fig~\ref{fig:determinant}.
This means that subtracting the lensing bias due to the source trispectrum will always cause only a small loss in lensing SNR.
Indeed, as shown in Fig~\ref{fig:determinant} (solid line), the lensing noise for the standard QE and the bias hardened estimators are very similar, meaning that subtracting the source bias causes a negligible increase in the lensing noise. This is consistent with the findings of \cite{Fabbian_2019}.

Finally, the non-Gaussianity of the point sources also contributes an extra noise term in the lensing reconstruction, which we haven't included.
However, this noise is only important when the bias to CMB lensing is also important. 
Thus, by making sure that the non-Gaussian point sources (and other foregrounds) do not significantly bias the lensing signal, we are making sure that their noise contribution is also small.

\section{Biases to CMB lensing \& foreground hardening}
\label{sec:biases}

\begin{figure}[!h]
\includegraphics[width=\linewidth]{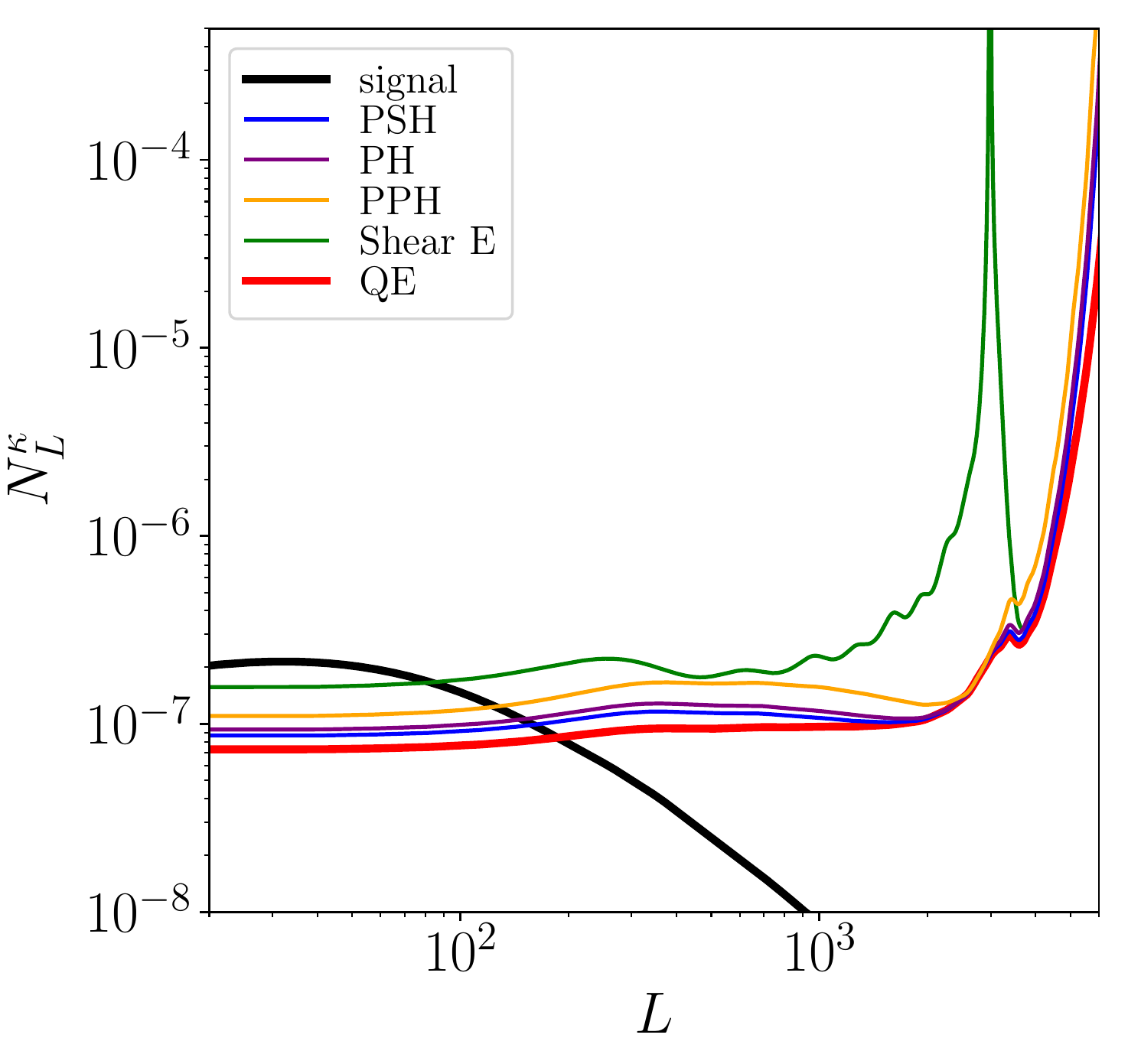}
\caption{
Noises of the Point Source Hardened (PSH), Profile Hardened (PH), Point source and Profile Hardened (PPH), Shear, and standard quadratic (QE) estimators for $\ell_\text{max,T} = 3500$. In black is the lensing signal $C^\kappa_L$. Hardening against a single source (PSH,PH) marginally increases the noise over QE, whereas hardening against multiple sources (PPH) results in a larger noise penalty. We also note that the noise cost of bias-hardening increases with the size of the source profile $u_{\bm{\ell}}$, which is discussed further in Appendix~\ref{app:noise price for larger profile}.}
\label{fig:noise}
\end{figure}

In this section we quantify the biases for three different bias-hardening schemes, and compare these results to the standard QE and Shear estimators. We consider two source components:  point-sources $f^\text{PS}_{\bm{\ell},\bm{L}-\bm{\ell}}=1$ and sources with tSZ-like profiles 
$f^\text{tSZ}_{\bm{\ell},\bm{L}-\bm{\ell}}
=
u_{\bm{\ell}}
u_{\bm{L}-\bm{\ell}}
/
u_{\bm{L}}
$. The profile $u_{\bm{\ell}}$ is calculated by taking the square root of the tSZ power spectrum, which is measured from the simulations. Our three different bias-hardening schemes harden against one or both of these source components. The Point Source Hardened (PSH) estimator hardenens against point sources; the Profile Hardened (PH) estimator hardens against a tSZ-like cluster profile, as described in App.~\ref{app:tsz profile}; and the Point source and Profile Hardened (PPH) estimator simultaneously hardens against both.
We show the noise power spectra for these various estimators in Fig.~\ref{fig:noise}, and performed various checks of our code in App.~\ref{app:pipeline_checks}.

\subsection{Bias to CMB lensing from non-Gaussian foregrounds}

The presence of a non-Gaussian foreground component $s$ in the temperature map produces three bias terms to CMB lensing, referred to as primary, secondary and trispectrum terms \cite{2019PhRvL.122r1301S, 2014JCAP...03..024O, 2018PhRvD..97b3512F, 2014ApJ...786...13V}.
Indeed, if $T = T^\text{CMB} + s$, then the quadratic estimator $\mathcal{Q}$ applied to $T$ can be expanded bilinearly as:
\beq
\mathcal{Q}[T, T]
=
\mathcal{Q}[T^\text{CMB}, T^\text{CMB}]
+ 2 \mathcal{Q}[T^\text{CMB}, s]
+\mathcal{Q}[s,s],
\eeq
assuming the quadratic estimator is symmetric (or has been symmetrized in its arguments).

In cross-correlation with a mass tracer $g$, this simply leads to the bispectrum bias:
\beq
\langle g \mathcal{Q} \rangle
= 
\underbrace{\langle g \kappa_\text{CMB} \rangle}
_\text{lensing signal}
+
\underbrace{\langle g \mathcal{Q}[s, s] \rangle}
_\text{bispectrum bias}.
\eeq

For the auto-correlation, more terms arise:
\beq
\bal
\langle \mathcal{Q} \mathcal{Q} \rangle
&=
\underbrace{\langle \kappa_\text{CMB} \kappa_\text{CMB} \rangle}
_\text{lensing signal}\\
&+
\underbrace{2\langle \mathcal{Q}[T^\text{CMB}, T^\text{CMB}] \mathcal{Q}[s, s] \rangle}
_\text{Primary bispectrum bias}\\
&+
\underbrace{4\langle \mathcal{Q}[T^\text{CMB}, s] \mathcal{Q}[T^\text{CMB}, s] \rangle}
_\text{Secondary bispectrum bias}\\
&+
\underbrace{\langle \mathcal{Q}[s, s] \mathcal{Q}[s, s] \rangle}
_\text{Trispectrum bias}.\\
\eal
\eeq
The primary and secondary bias terms are due to the non-Gaussianity of the foreground and its correlation with the true CMB lensing signal. They are integrals of the $\kappa_\text{CMB} s s$ bispectrum. In particular, predicting them requires knowing the redshift distribution of the foreground.
On the other hand, the trispectrum bias is present whether or not the sources are correlated with CMB lensing, and only depend on the statistics (the trispectrum) of the projected 2d foreground map.

In what follows, we compute these biases for the various quadratic lensing estimators of interest. We follow the approach of \cite{2019PhRvL.122r1301S}, using the simulations from \cite{2010ApJ...709..920S}.

\subsection{Effectiveness of foreground hardening: expectations}

Here we explain which of these bias terms are nulled with bias-hardening.
The only knowledge of foregrounds used in the bias hardening process is the shape of the foreground bispectrum and power spectrum, through $f^s$. In particular, the procedure does not assume anything about the $ss \kappa_\text{CMB}$ bispectrum, which causes the primary and secondary biases, nor about the $ssss$ trispectrum, which causes the trispectrum bias.
Why would bias hardening help reducing foreground biases at all then?
The answer is somewhat subtle.

The bias hardening procedure nulls the linear response of the lensing estimator to the foreground $s$, i.e. the bispectrum
$\langle \mathcal{Q}^\text{BH} [s, s] \;s \rangle$.
This is very similar but not identical to the primary bispectrum bias
$\langle \mathcal{Q}^\text{BH} [s, s] \;\kappa_\text{CMB} \rangle$.
The two bispectra are all the more similar as the foreground map and the true lensing map are highly correlated.
Thus in principle, the primary bispectrum bias is only partially reduced by the bias-hardening.
In practice though, as we show below with simulated foregrounds, this reduction is very large.

Although the secondary contraction is also a bispectrum of the form $ s s \kappa_\text{CMB}$, it comes from a different contraction, $\langle \mathcal{Q}[T, s] \mathcal{Q}[T, s] \rangle$, which the bias hardened estimator was not designed to null. For this reason, we do not expect bias hardening to reduce the secondary contraction.
Indeed, we show below that the various lensing estimators have a similar level of secondary bias.

The case of the trispectrum bias is interesting.
In general, if $\mathcal{T}_{\bm{\ell}_1, \bm{\ell}_2, \bm{\ell}_3, \bm{\ell}_4}$ is the source trispectrum, the trispectrum bias to lensing is given by
\beq
\bal
&\langle \mathcal{Q}[s, s] \mathcal{Q}[s, s] \rangle_c
=\\
&N^2_{\boldsymbol{L}} 
\int
\frac{d^2\boldsymbol{\ell}}{(2\pi)^2} 
F_{\boldsymbol{\ell},\boldsymbol{L}-\boldsymbol{\ell}}
\int
\frac{d^2\boldsymbol{\ell '}}{(2\pi)^2} 
F_{\boldsymbol{\ell '}, -\boldsymbol{L}-\boldsymbol{\ell '}}
\mathcal{T}_{\bm{\ell}, \bm{L}-\bm{\ell}, \bm{\ell} ', -\bm{L} - \bm{\ell} '}.
\eal
\eeq
If the lensing weights of the bias-hardened estimator are denoted as 
$F^\text{BH}_{\boldsymbol{\ell},\boldsymbol{L}-\boldsymbol{\ell}}$,
then the zero linear response to the foreground can be translated as:
\beq
\int 
\frac{d^2\boldsymbol{\ell}}{(2\pi)^2} 
F^{\kappa^\text{BH}}_{\boldsymbol{\ell},\boldsymbol{L}-\boldsymbol{\ell}}
f^s_{\bm{\ell},\bm{L}-\bm{\ell}}
=0.
\label{eq:no response to sources}
\eeq
This does not null the trispectrum bias in general.
However, it does in the case of Poisson sources with identical profiles.
Indeed, in this case, the trispectrum $\mathcal{T}_{\bm{\ell}_1, \bm{\ell}_2, \bm{\ell}_3, \bm{\ell}_4} = \bar{n} \langle s^4 \rangle u_{\bm{\ell}_1} u_{\bm{\ell}_2} u_{\bm{\ell}_3} u_{\bm{\ell}_4} $ is separable, such that the trispectrum bias becomes:
\beq
\bar{n}\langle s^4 \rangle
N^2_{\boldsymbol{L}} 
\left[
\int
\frac{d^2\boldsymbol{\ell}}{(2\pi)^2} 
F^\text{BH}_{\boldsymbol{\ell},\boldsymbol{L}-\boldsymbol{\ell}}
u_{\bm{\ell}}
u_{\bm{L}-\bm{\ell}}
\right]^2.
\label{eq:bias from trispectrum}
\eeq
Since the response $f^s_{\bm{\ell},\bm{L}-\bm{\ell}} = u_{\bm{\ell}} u_{\bm{L}-\bm{\ell}}/u_{\bm{L}}\propto u_{\bm{\ell}} u_{\bm{L}-\bm{\ell}}$, we see that Eq.~\eqref{eq:no response to sources} implies that the trispectrum bias is exactly nulled. 
However, it is no longer generally true if the sources are clustered, or if there is a distribution of profile sizes or shapes.
Because our profile hardening assumes a single fiducial source profile, the trispectrum bias is also not exactly nulled when the foreground sources have a range of profiles, as is the case for the tSZ.

In summary, the PH estimator should significantly reduce the tSZ trispectrum bias, while the PSH estimator should exactly null the point source trispectrum. Both should largely reduce the primary bias, with no \textit{a priori} expectation for the secondary bias. 

For bias hardening against more complex foregrounds, such as clustered sources, the effectiveness of this simple bias hardening is not guaranteed. However, if the statistical properties of the source distribution is known, it is nonetheless possible to extend our formalism to harden against non-linear clustering. This has been explored in the context of line intensity mapping in \cite{2018JCAP...07..046F}, and we believe that a similar treatment could be helpful for mitigating biases to CMB lensing.
In what follows, we quantify this effect precisely, by running the hardened estimators on foreground simulations with a realistic clustering.

\subsection{Effectiveness of foreground hardening: simulations}

To quantity the biases of each estimator for each of the extragalactic foregrounds, we use simulated maps \cite{2010ApJ...709..920S} of the CIB, tSZ, kSZ, radio point sources, and $\kappa_\text{CMB}$ at 148 GHz. We rescale each of the foregrounds (0.38 for CIB,
0.7 for tSZ, 0.82 for kSZ, 1.1 for radio PS) to match \cite{2013JCAP...07..025D}, then subtract the mean from each foreground, and finally mask each foreground using a matched filter for point sources with a flux cut of 5 mJy and mask patch radius 3'. The resulting power spectra are plotted in Appendix~\ref{app:foreground power spectra}. 

For CMB measurements, we consider an experiment similar to Simons Observatory \cite{2019JCAP...02..056A}, with a white noise level of 6 $\mu$K-arcmin and a Gaussian beam with full-width at half maximum of $1.4'$.

\begin{figure}[!h]
\centering
\includegraphics[width=\linewidth]{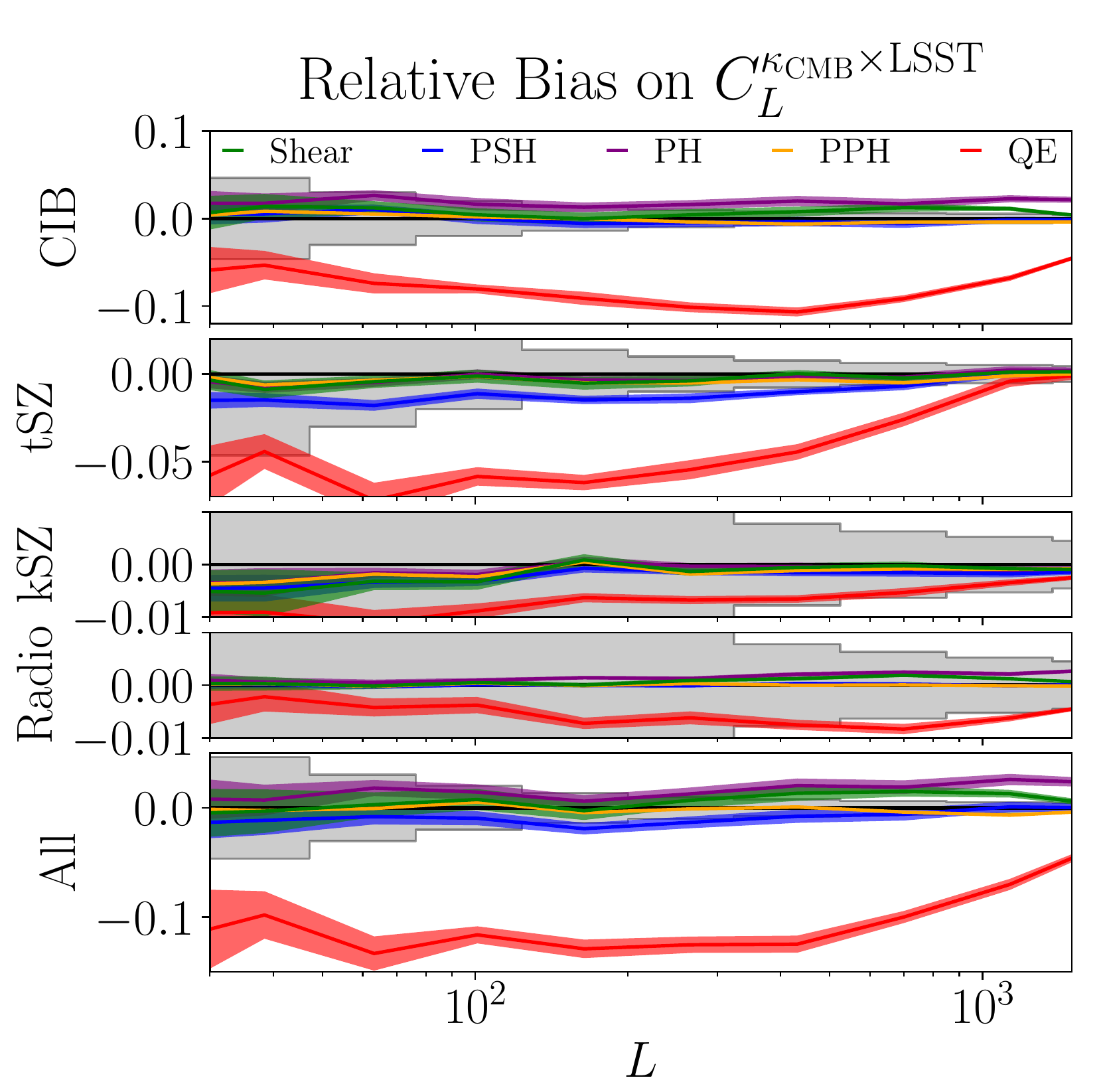}
\caption{Relative bias on the cross with LSST at $\ell_\text{max,T}=3500.$ The gray boxes are the statistical uncertainties when reconstructing $C^{\kappa\delta_g}_L$ with the standard QE. We find that hardening against point sources (PSH) nulls or dramatically reduces the bias from radio point sources and CIB, whereas hardening against a tSZ-like profile (PH) nulls or significantly reduces the bias from tSZ and kSZ. Hardening against both (PPH) effectively nulls the bias from all foregrounds, resulting in a bias that is roughly an order of magnitude smaller than the bias to the standard QE.}
\label{fig:cross-correlation biases}
\end{figure}

\begin{figure}[!h]
\centering
\includegraphics[width=.95\linewidth]{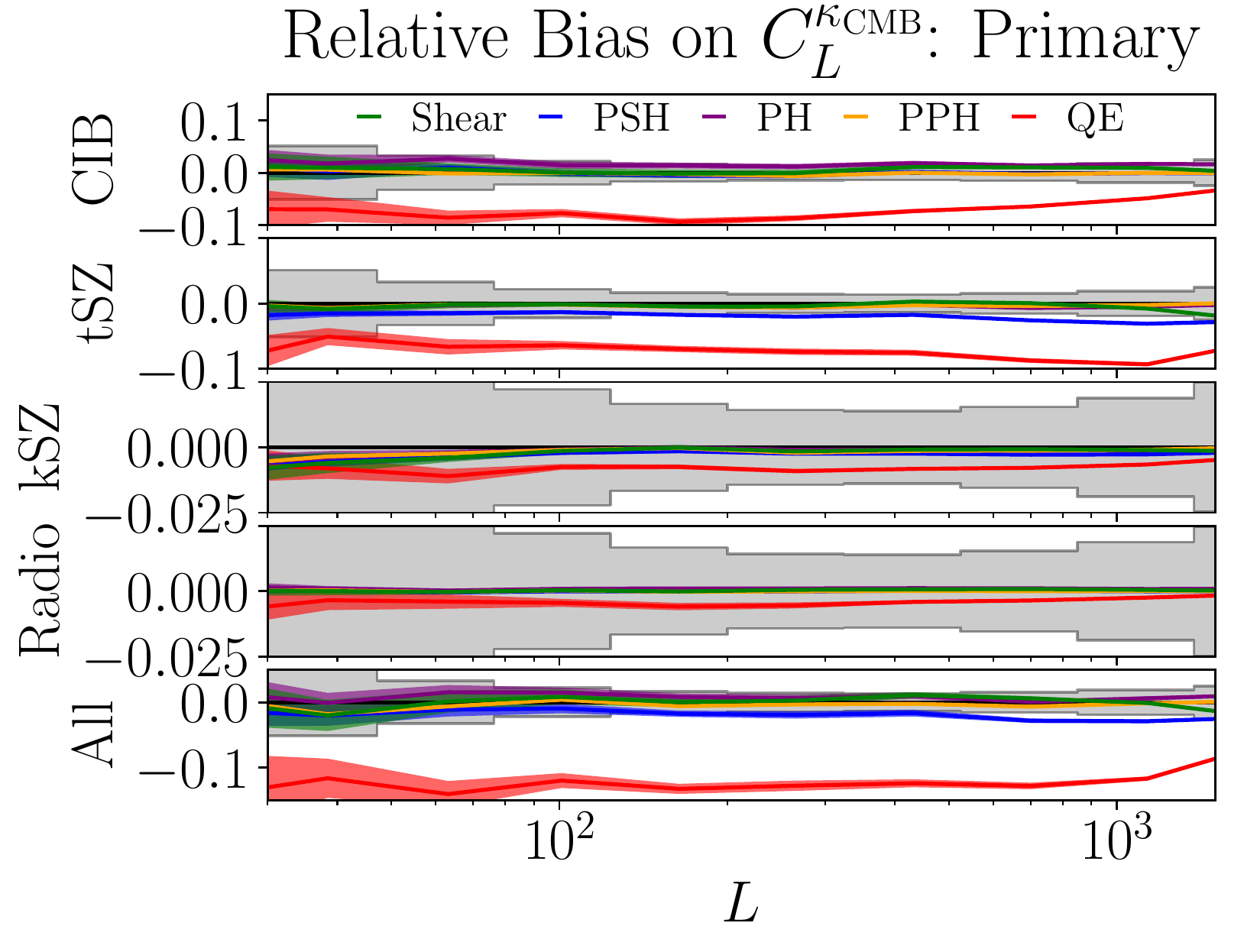}
\includegraphics[width=.95\linewidth]{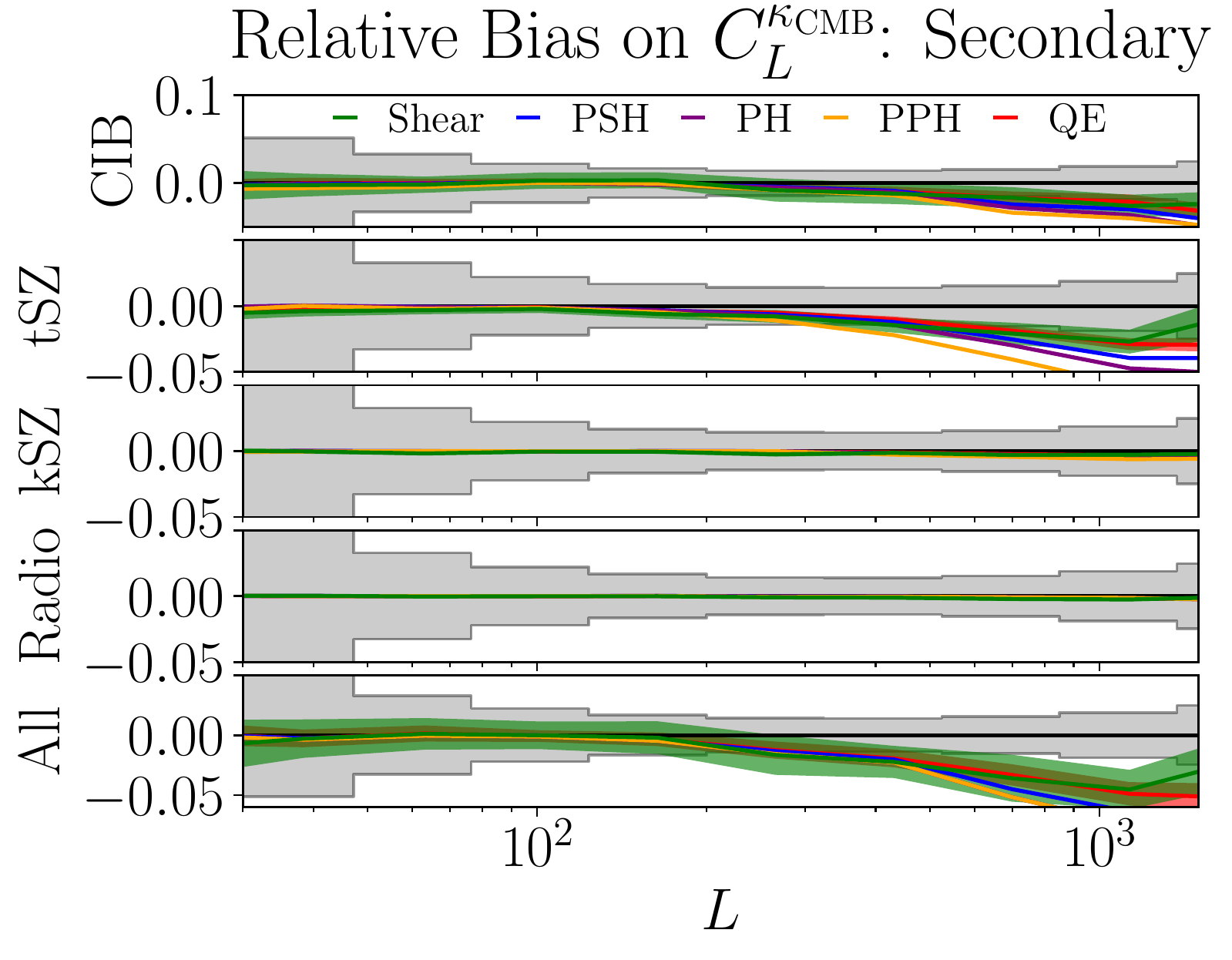}
\includegraphics[width=.95\linewidth]{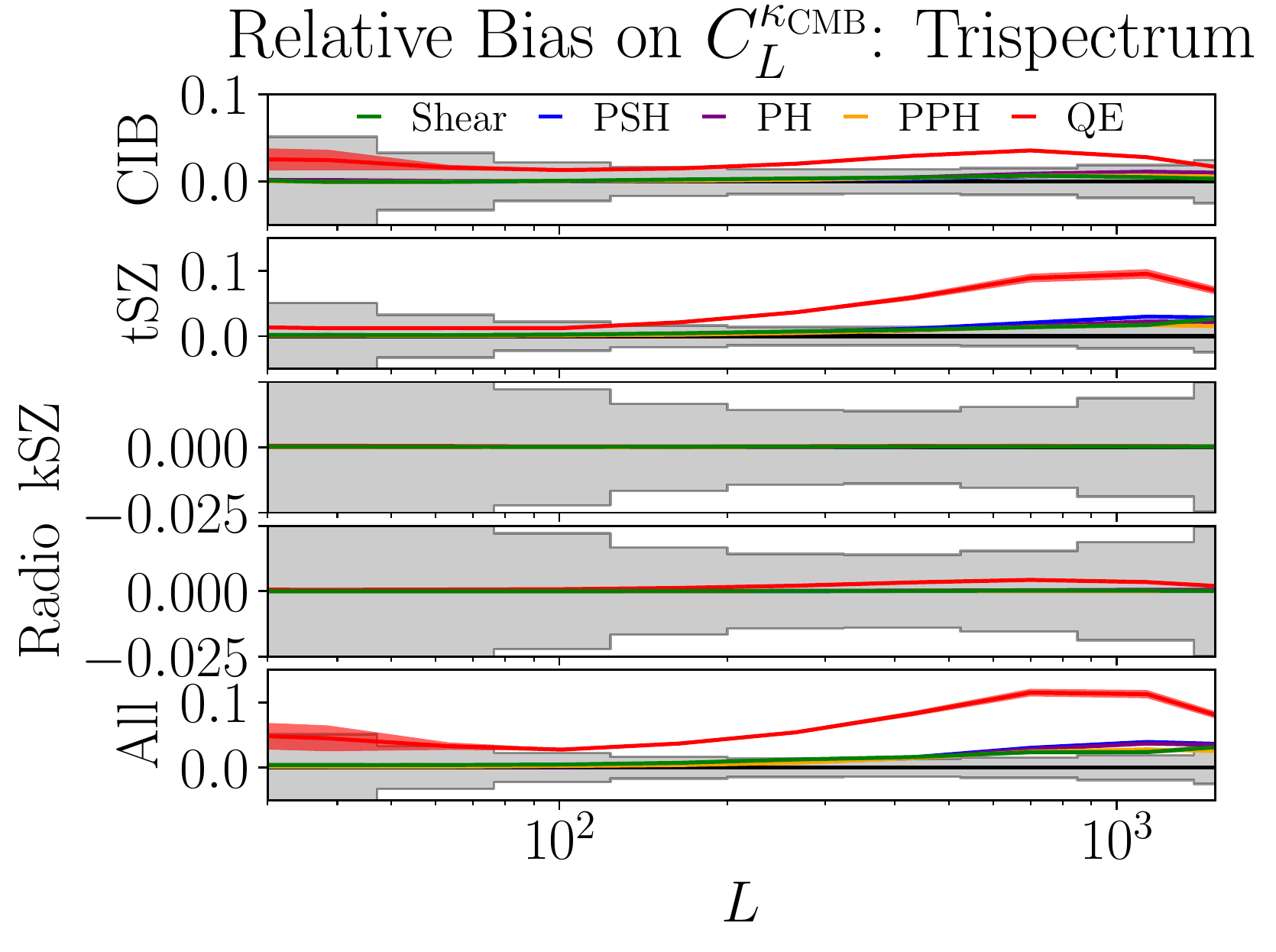}
\caption{Bias to the lensing auto-correlation at $\ell_\text{max,T} = 3500$. The gray boxes are the statistical uncertainties when reconstructing $C^\kappa_L$ with the standard QE. Hardening against point sources (PSH) significantly reduces the primary and trispectrum bias from radio point sources and CIB, whereas hardening against a tSZ-like profile significantly reduces these biases from tSZ. Simultaneously hardening against both (PPH) effectively nulls the primary bias. Bias-hardening has no significant effect on the secondary bias.}
\label{fig:auto-correlation biases}
\end{figure}

As in \cite{2019PhRvL.122r1301S}, we re-weight the halos from \cite{2010ApJ...709..920S} to match the redshift distribution of the LSST gold sample ($dn/dz \propto (z/0.24)^2 e^{-z/0.24}/0.48$) \cite{2009arXiv0912.0201L} and obtain a projected galaxy number density count $\delta_g$. Altogether, a set of these maps (foregrounds, $\kappa_\text{CMB}$, and $\delta_g$) allow us to calculate the biases to the cross-correlation of CMB lensing with galaxy number density counts $C^{\kappa\delta_g}_L$, as well as the primary, secondary, and trispectrum biases to the lensing auto-correlation $C^{\kappa\kappa}_L$. 
We run the estimators on 81 flat square cutouts from these maps, obtaining the biases plotted in Fig.~\ref{fig:cross-correlation biases} and \ref{fig:auto-correlation biases}.

For the cross-correlation of CMB lensing with galaxy counts (Fig.~\ref{fig:cross-correlation biases}), we find that the standard minimum variance quadratic estimator acquires a $\sim 10$\% bias from the CIB and tSZ for reasonable values of $\ell_\text{max,T}$. Hardening against a single source component with a tSZ-like profile (PH) effectively nulls the bias from both tSZ and kSZ. Hardening against just point sources (PSH) is complementary, effectively nulling the radio point source and CIB biases. When simultaneously hardening against both tSZ-like profiles and point sources (PPH), we're able to null the biases from all extragalactic foregrounds, creating a highly robust estimator of the cross-correlation. 

For the CMB lensing auto-correlation we again find that the standard QE acquires a $\sim 10$\% bias, primarily due to the CIB and tSZ. This is consistent with the results found in \cite{2014JCAP...03..024O}. In Fig.~\ref{fig:auto-correlation biases} we see that the PH estimator nulls the primary bias from tSZ, and significantly reduces the remaining primary and trispectrum biases. Likewise the PSH estimator nulls the primary bias from point sources, while also significantly reducing remaining primary and trispectrum biases. The PPH estimator essentially nulls all primary biases, and reduces the trispectrum bias to a level comparable to Shear. As discussed in the previous section, there is no reason to expect a significant reduction to the secondary biases from bias-hardening. In practice we find that the secondary bias to bias-hardened estimators is slightly worse than (but comparable to) the bias to the standard QE. We note that the secondary bias is negative, whereas the bias from the trispectrum is positive. This results in a cancellation in the overall bias to the lensing power, allowing higher values of $\ell_\text{max,T}$ than one might expect from looking at the secondary and trispectrum biases alone. 

\begin{figure}[!h]
\centering
\includegraphics[width=\linewidth]{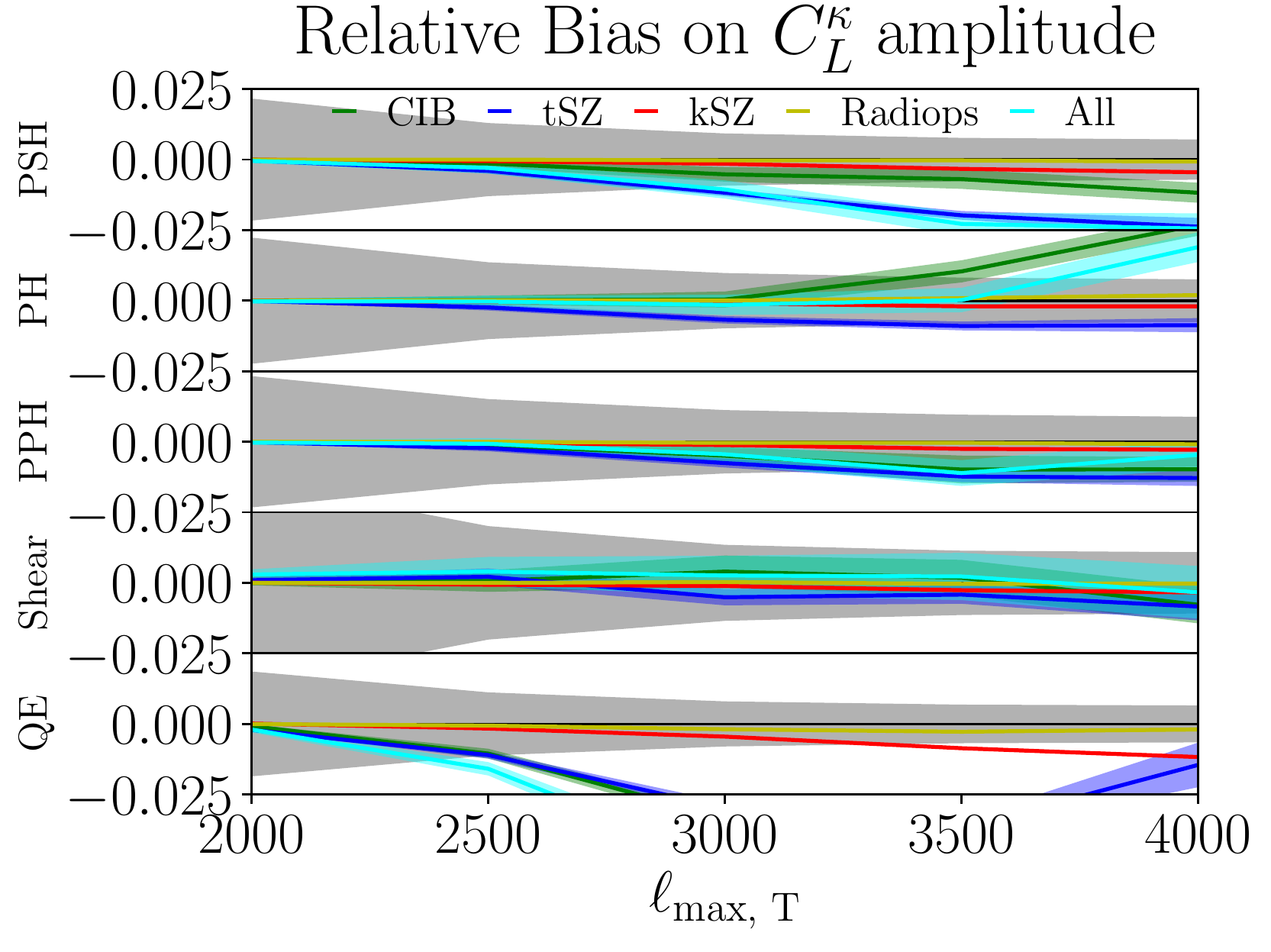}
\includegraphics[width=\linewidth]{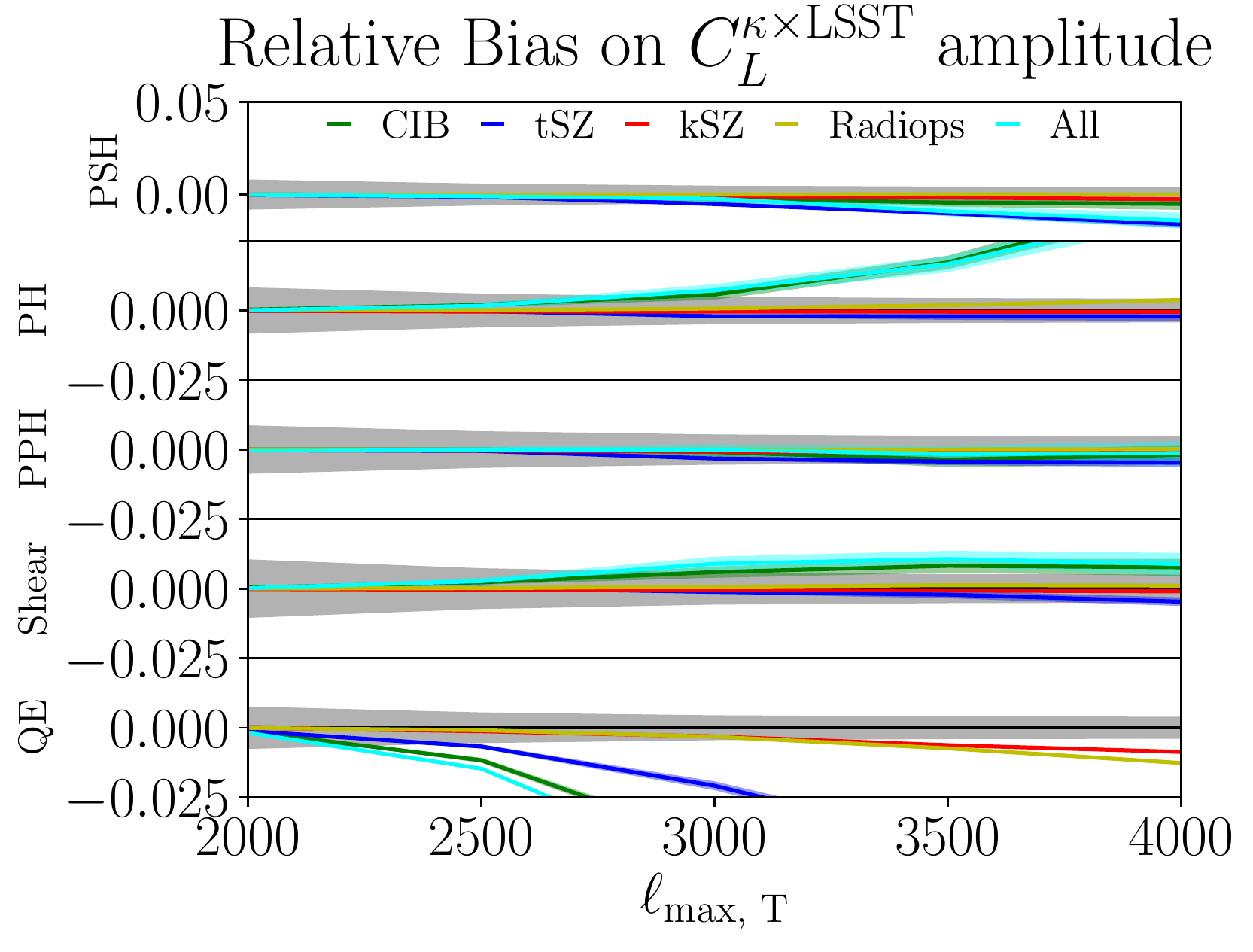}
\caption{Relative bias on the lensing amplitude and the cross-correlation with LSST as a function of $\ell_\text{max, T}$. In gray is the $1\sigma$ error for each estimator $1/\text{SNR}(A_\text{lens})$. CIB and tSZ are the main sources of bias to the standard QE. The Point Source Hardened (PSH) estimator drastically reduces the bias from CIB, whereas the Profile Hardened (PH) estimator reduces the bias from tSZ. The Point source and Profile Hardened (PPH) estimator reduces the bias from both tSZ and CIB to a level comparable to Shear while at a lower noise cost.
}
\label{fig:bias-to-amplitude}
\end{figure}

As a figure of merit for the performance of each estimator, we use the signal-to-noise ratio (SNR) of the lensing amplitude $A_\text{lens}$, defined for each $L$ as $C^{\kappa,m}_L/C^\kappa_L$, where $C^{\kappa,m}_L$ and $C^\kappa_L$ are the measured and true lensing power spectra respectively. By inverse variance weighting the measurement of $A_\text{lens}$ for each $L$, we obtain an estimator for the lensing amplitude:
\beq
\bal
\hat{A}_\text{lens}
=
\int
\frac{d^2\bm{L}}{(2\pi)^2}
\frac{C^{\kappa,m}_L}{C^\kappa_L}
\frac{(C^\kappa_L)^2}{\sigma^2_L}
\bigg/
\int
\frac{d^2\bm{L}}{(2\pi)^2}
\frac{(C^\kappa_L)^2}{\sigma^2_L}
\\
\text{with }
\sigma^2_L = 2(C^\kappa_L + N_L)^2.
\eal
\eeq
Since $A_\text{lens}$ has a fiducial value of $1$, the SNR and bias of $A_\text{lens}$ are (assuming $f_\text{sky}=1$)
\beq
\bal
\text{SNR}^2(A_\text{lens})
&=
4\pi
\int
\frac{d^2\bm{L}}{(2\pi)^2}
\frac{(C^\kappa_L)^2}{\sigma^2_L}
\\
\text{bias}(A_\text{lens})
&=
\int
\frac{d^2\bm{L}}{(2\pi)^2}
\frac{\text{bias}(C^\kappa_L)}{C^\kappa_L}
\frac{(C^\kappa_L)^2}{\sigma^2_L}
\bigg/
\int
\frac{d^2\bm{L}}{(2\pi)^2}
\frac{(C^\kappa_L)^2}{\sigma^2_L}.
\eal
\eeq
Similar definitions for the amplitude of the CMB-LSST cross-correlation can be made by replacing the lensing power $C^\kappa_L$ with $C^{\kappa\delta_g}_L$ and by setting $\sigma^2_L = (C^\kappa_L + N^\kappa_L)(C^{\delta_g}_L+N^{\delta_g}_L) + (C^{\kappa\delta_g}_L)^2$ in the equations above.
We show the best SNR achievable in auto and cross-correlation for the various estimators, while maintaining the foreground bias below $1\sigma$, in Tab.~\ref{tab:best_snr_comparison}.

\begin{figure}[!h]
\centering
\includegraphics[width=\linewidth]{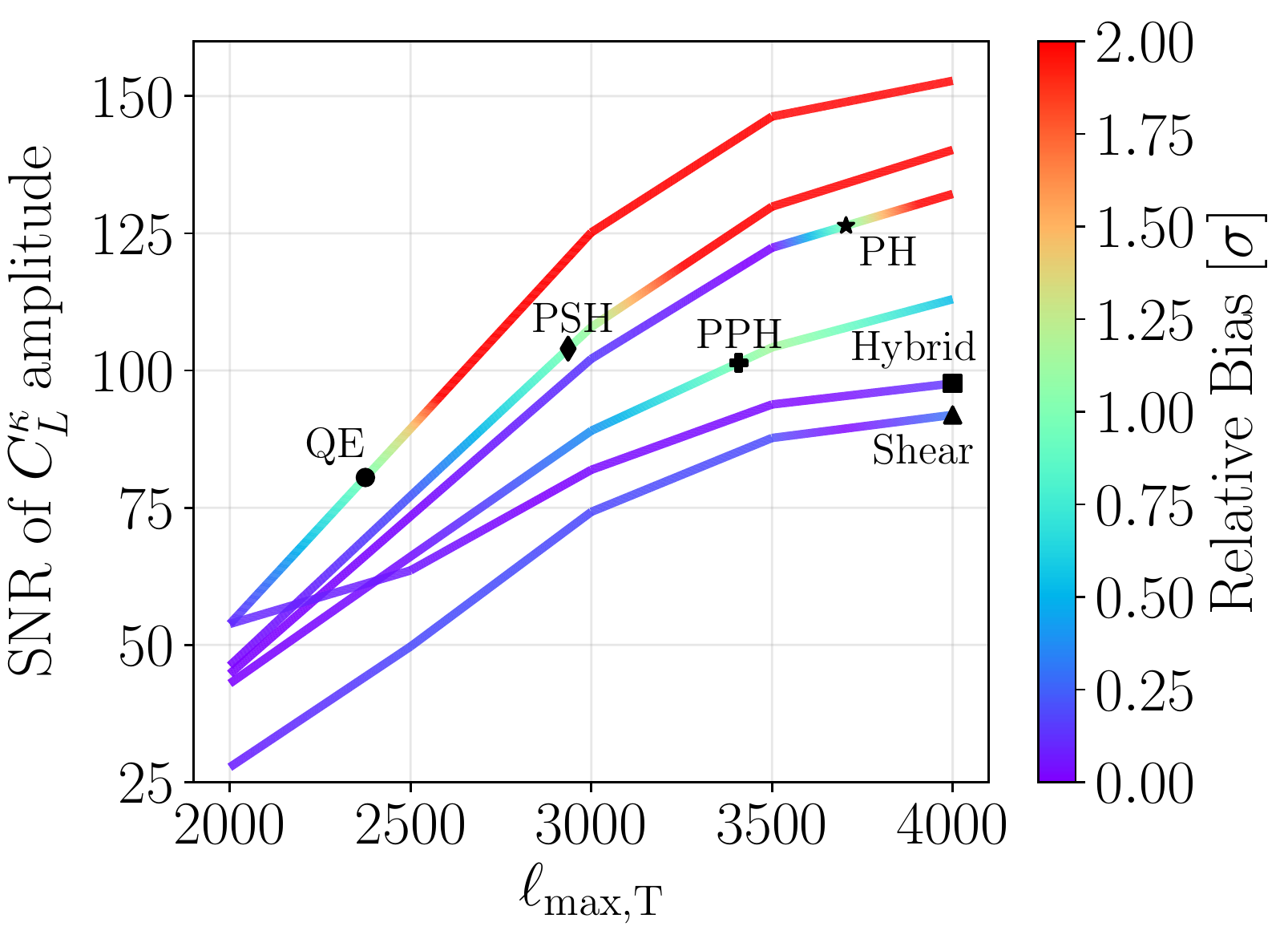}
\includegraphics[width=\linewidth]{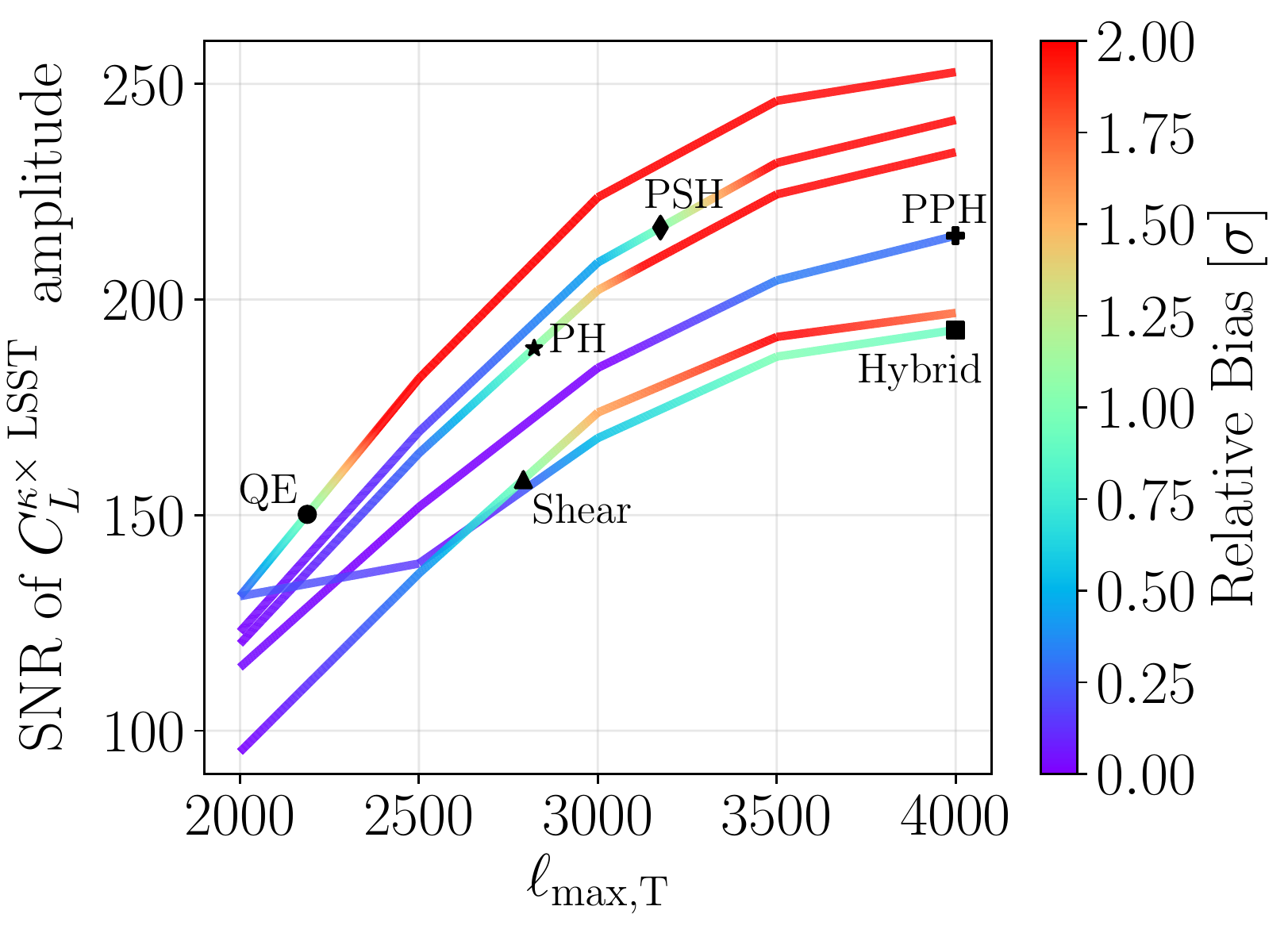}
\caption{Total signal-to-noise of the lensing amplitude (top) and the cross correlation with LSST (bottom). The black markers denote the highest $\ell_\text{max, T}$ where the bias is less than $1\sigma$. We find that the Profile Hardened (PH) estimator reconstructs the lensing amplitude with the highest signal to noise: a $\sim50$\% improvement over the standard QE. We find that the Point Source Hardened (PSH) and Point source and Profile Hardened (PPH) reconstruct the cross with similar SNRs: a $\sim50$\% improvement over the standard QE. The Hybrid estimator is a combination of Shear and QE, and is described in greater detail in Appendix~\ref{app:hybrid estimator}. We note that the color-scale saturates at $2\sigma$, and that the bias to the standard QE can be $>10\sigma$ at $\ell_\text{max,T}\gtrsim 3000$.
}
\label{fig:SNR of auto and cross}
\end{figure}

For the auto-correlation, we find that the Profile Hardened (PH) estimator achieves the highest SNR before $\text{bias}(A_\text{lens})$ becomes larger than the noise $\sigma=1/\text{SNR}(A_\text{lens})$, as shown in Fig.~\ref{fig:SNR of auto and cross}. This is a $\sim 50$\% improvement over the standard QE. The PH estimator's $1\sigma$ bias threshold occurs at $\ell_\text{max,T}\sim3700$. From Fig.~\ref{fig:bias-to-amplitude}, we see that this high value of $\ell_\text{max,T}$ is due to a cancellation in the tSZ and CIB biases. A more conservative stopping point would be when any one of the foregrounds crosses the $1\sigma$ threshold, which would be around $\ell_\text{max,T}\sim3200$ for the PH estimator. We note that even at this lower $\ell_\text{max,T}$, the PH estimator still has the highest SNR and is still a $\sim 30$\% improvement over QE, which crosses the $1\sigma$ bias threshold at $\ell_\text{max,T}\sim 2400$.

For the cross-correlation, the PSH and PPH achieve similar SNRs at the $1\sigma$ bias threshold, improving upon QE by $\sim 40$\%. We note that the PPH has a stable bias across the entire range of realistic $\ell_\text{max,T}$'s, making it a conservative yet competitive option for reconstructing the cross-correlation.

We conclude by noting that for an experiment with $\ell_\text{max,T} \sim 3000$, the PH and PPH estimators reconstruct the auto- and cross-correlations respectively with the highest SNR while achieving a sub-percent level bias, as shown in Tab.~\ref{tab:snr_lmax3000}.
\begin{table}[!h]
\begin{tabular}{c|c|c}
      &  SNR in auto $(\ell_\text{max,T})$ & SNR in cross $(\ell_\text{max,T})$\\
     \hline
     \hline
     PSH & 104 (2936) & 217 (3175)\\
     \hline
     PH &  126 (3705)  & 189 (2822)\\
     \hline
     PPH & 101 (3408) & 215 (4000)\\
     \hline
     Shear & 92 (4000) & 158 (2792)\\
     \hline
     QE & 81 (2375) & 150 (2188)
\end{tabular}
\caption{
Comparison of the total signal-to-noise ratio on the amplitude of the CMB lensing auto-spectrum (first column) and cross-spectrum with LSST-like galaxies (second column).
In all cases, the maximum temperature multipole $\ell_\text{max, T}$ is selected to guarantee a foreground bias smaller than $1\sigma$.
}
\label{tab:best_snr_comparison}
\end{table}

\begin{table}[!h]
\begin{tabular}{c|c|c|c}
      &  
      $\substack{\text{Rel. bias to}\\ \text{auto (cross) [$\%$]} }$ & 
      $\substack{\text{Rel. bias to}\\ \text{auto (cross) [$\sigma$]} }$ & 
      \text{SNR on $C^\kappa_L$ ($C^{\kappa\times \text{LSST}}_L$)}\\
      \hline
      \hline
      PSH & $-1.1$ ($-0.2$) & $-1.14$ ($-0.46$) & 108 (209)\\
      \hline
      PH & $-0.1$ (0.7) & $-0.15$ (1.47) & 102 (202)\\
      \hline
      PPH & $-0.4$ (0.1) & $-0.39$ (0.01) & 89 (184)\\
      \hline
      Shear & 0.3 (0.9) & 0.19 (1.55) & 74 (173)\\
      \hline
      QE & $-4.9$ ($-5.0$) & $-6.15$ ($-11.16$) & 125 (223)
\end{tabular}
\caption{
Relative bias (in both \% and $\sigma$ units) and SNR for each estimator at $\ell_\text{max,T} =3000$. The PH and PPH estimators reconstruct the auto- and cross-correlations respectively with the highest SNR while achieving a sub-percent level bias.
}
\label{tab:snr_lmax3000}
\end{table}

\section{Discussion and conclusions}
\label{sec:conclusions}

In this study we revisited foreground-hardened lensing quadratic estimators,
described how to harden against arbitrary profiles, and extended the formalism to
simultaneously harden against an arbitrary number of extragalactic foregrounds.
We showed that the point source hardened estimator (PSH) reduces not only the bias from radio point sources, but also from the cosmic infrared background, the thermal and the kinematic Sunyaev-Zel'dovich effects.

Motivated by the extended nature of tSZ clusters, we modify the point source hardening technique to deproject the lensing bias from uncorrelated clusters (PH estimator), or simultaneously point sources and extended clusters (PPH estimator). We propose a simple and sufficient method to approximate the input cluster profile from the data itself, by considering the observed tSZ power spectrum from the same data.

The signal-to-noise cost from deprojecting the point source bias is small ($\sim 15\%$ for Point Source Hardening), and is more than compensated by enabling the use of higher multipoles: overall, the SNR on the lensing power spectrum is increased by $\sim$ 56\% when using Profile Hardening (PH), while for cross-correlations, both PSH and PPH provide a $\sim 45\%$ increase in SNR compared to QE. Part of the improvement of PH and PPH estimators comes from the increase in the maximum multipole that can be included in the lensing reconstruction.

The PSH, PH and PPH estimators improve the lensing SNR over the shear and hybrid shear estimators of \cite{2019PhRvL.122r1301S}, although at the cost of a slightly larger bias in cross-correlation.
They also provide consistency checks for the standard quadratic estimator, having a smaller foreground bias for a given $\ell_\text{max}$ in temperature.
Overall, these estimators outperform the standard lensing quadratic estimator both in auto and cross-correlation, both in terms of the lensing SNR and in terms of the foreground bias. We recommend their use in the CMB lensing analyses of upcoming temperature-dominated CMB data such as the Simons Observatory. Future surveys that are dominated by polarization reconstruction such as CMB-S4 \cite{2019arXiv190704473A}, will still benefit from improved robustness, since temperature reconstruction will contribute a non-trivial fraction to the SNR. We note that a realistic analysis will require modeling of the higher order biases $N^{(i)}$ \cite{Fabbian_2019} that appear in lensing reconstruction and will likely be non-negligible for the next generation of surveys.

Finally, foreground hardening can and should be implemented in conjunction with other methods of foreground mitigation, such as masking \cite{2014ApJ...786...13V} and multi-frequency cleaning in one or both legs of the estimator.
We shall report on the optimal combination of these methods in an upcoming paper. 

\section*{Acknowledgments}

We thank Anthony Challinor, Colin Hill, Antony Lewis, Mathew Madhavacheril, Blake Sherwin, Omar Darwish, Alex van Engelen, Neelima Sehgal, David Spergel, Martin White and the participants of the Sussex CMB lensing workshop for all their feedback. We give a special thanks to Zoom for facilitating this collaboration throughout the COVID-19 pandemic. E.S. is supported by the Chamberlain fellowship at Lawrence Berkeley National Laboratory. S.F. is supported by the Physics Division of Lawrence Berkeley National Laboratory.
This work used resources of the National Energy Research Scientific Computing Center, a DOE Office of Science User Facility supported by the Office of Science of the U.S. Department of Energy under Contract No. DE-AC02-05CH11231.

\clearpage
\onecolumngrid
\appendix

\section{Foreground power spectra}
\label{app:foreground power spectra}
\begin{figure}[!h]
\centering
\includegraphics[width=.5\linewidth]{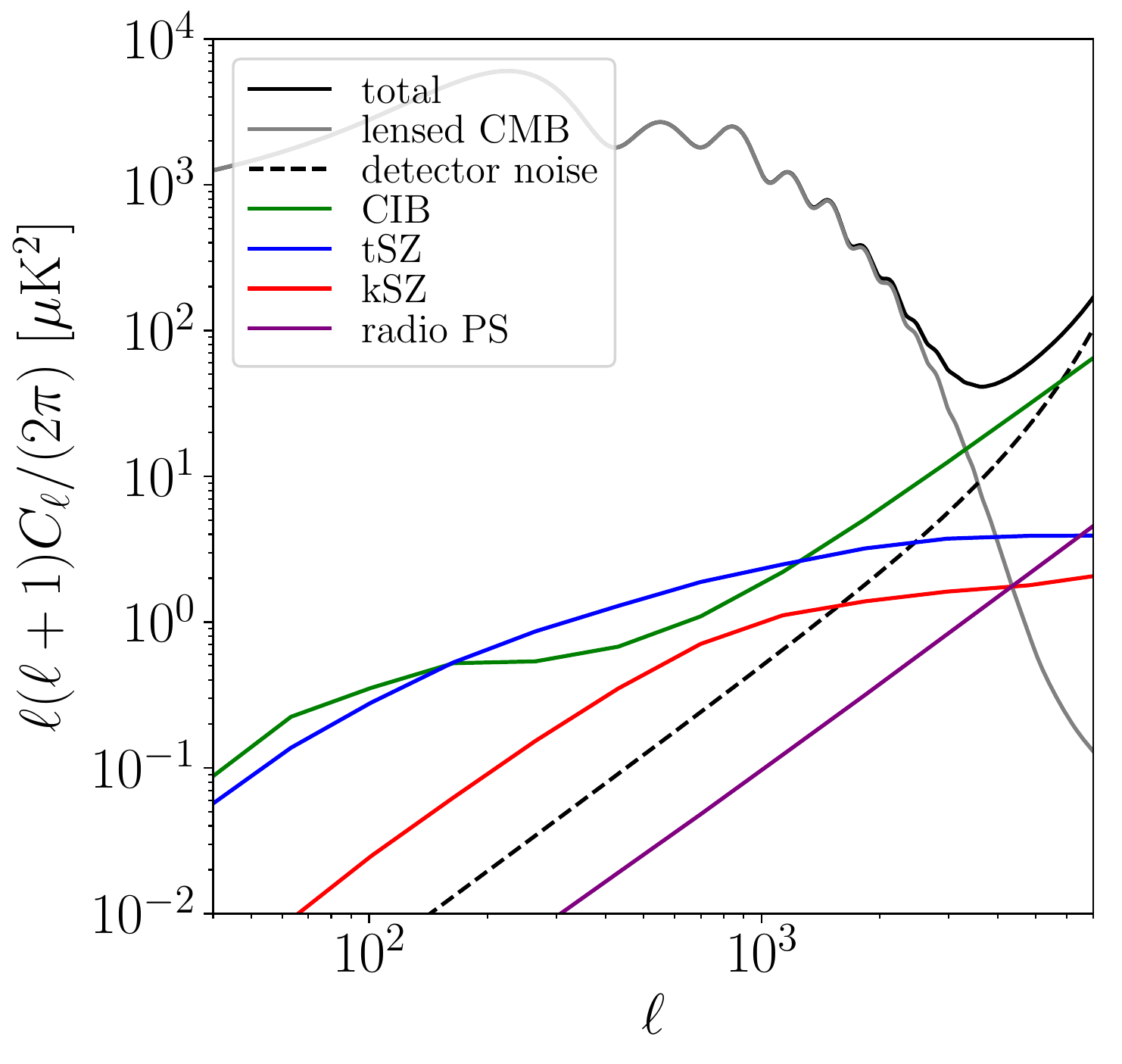}
\caption{Foreground power spectra at 148 GHz after being masked with a match-filter with a flux cut of 5 mJy and a mask patch radius of 3'. We assume a 6 $\mu$K-arcmin noise, and a 1.4' beam.}
\label{fig:foreground_power}
\end{figure}

Our processing of the Sehgal simulations follows \cite{2019PhRvL.122r1301S}.
After masking the sources detected in the foreground maps at 5 mJy, which are detected at about 5$\sigma$ given the noise level assumed, the power spectrum of each foreground at 148 GHz is shown in Fig.~\ref{fig:foreground_power}.
\section{Hybrid estimator}
\label{app:hybrid estimator}
The hybrid estimator is defined to be 
\begin{equation}
    \hat{\kappa}_L^\text{H} = N_L^\text{H}\left(\hat{\kappa}_L^\text{QE}/N_L^\text{QE} + \hat{\kappa}_L^\text{S} / N_L^\text{S}\right)
\end{equation}
where the standard estimator $\hat{\kappa}_L^\text{QE}$ is calculated using $\ell_\text{min,T} \leq \ell \leq 2000$ and the shear estimator $\hat{\kappa}_L^\text{S}$ is calculated using $2000 \leq \ell \leq \ell_\text{max,T}$. The noise of the hybrid estimator is simply 
$$N_L^\text{H} = \frac{1}{1/N_L^\text{QE} + 1/N_L^\text{S}}.$$ 
assuming that shear and QE are uncorrelated. The bias $B_L$ to the lensing signal is defined by
\begin{align}
    C^\text{QE}_L &= C_L^\kappa + N_L^\text{QE} + B_L^\text{QE}\\
    C^\text{S}_L &= C_L^\kappa + N_L^\text{S} + B_L^\text{S}.
\end{align}
Assuming that the noises of $\hat{\kappa}_L^\text{QE}$ and $\hat{\kappa}_L^\text{S}$ are uncorrelated, the bias to the Hybrid estimator is
\begin{equation}
B_L^\text{H} =  
\left(N_L^\text{H}\right)^2\left(\frac{B_L^\text{QE}}{\left(N^\text{QE}\right)^2} + \frac{B_L^\text{S}}{\left(N^\text{S}\right)^2} \right).
\end{equation}
Here $B_L^\text{QE}$ is the bias when the standard quadratic estimator is calculated using $\ell_\text{min,T} \leq \ell \leq 2000$. We approximate $B_L^\text{S}$ as the bias from Shear when calculated with $\ell_\text{min,T} \leq \ell \leq \ell_\text{max,T}$. This approximation should be OK since the contribution from $\ell_\text{min,T}\leq \ell\leq2000$ is negligible. The bias to the hybrid estimator's cross-correlation with LSST is calculated by inverse noise weighting the biases to QE and Shear.
\section{Derivation of Eq. \eqref{eq:general_noise}}
\label{app:derivation of general noise}
We start with
\beq
\bm{M}_{\bm{L}}
=
\begin{pmatrix}
1 & 
N^\kappa_{\boldsymbol{L}}
\mathcal{R}^{\kappa,s_1}_{\boldsymbol{L}} &
\cdots &
N^\kappa_{\boldsymbol{L}}
\mathcal{R}^{\kappa,s_n}_{\boldsymbol{L}}\\
N^{s_1}_{\boldsymbol{L}}
\mathcal{R}^{s_1,\kappa}_{\boldsymbol{L}} & 
1 &
\cdots
&
N^{s_1}_{\boldsymbol{L}}
\mathcal{R}^{s_1,s_n}_{\boldsymbol{L}}
\\
\vdots &
\vdots
&
\vdots
&
\vdots\\
N^{s_n}_{\boldsymbol{L}}
\mathcal{R}^{s_n,\kappa}_{\boldsymbol{L}}
&
N^{s_n}_{\boldsymbol{L}}
\mathcal{R}^{s_n,s_1}_{\boldsymbol{L}}
&
\cdots
&
1
\end{pmatrix}
\equiv
\begin{pmatrix}
1 & \bm{w}^T_{\bm{L}} \\
\bm{v}_{\bm{L}} & \bm{A}_{\bm{L}}
\end{pmatrix},
\eeq
where $\bm{A}_{\bm{L}}$ is a $n\times n$ matrix and $\bm{w}_{\bm{L}},\bm{v}_{\bm{L}}$ are $n$-dimensional column vectors. By splitting the matrix into blocks, we can express its inverse and determinant as
\beq
\bm{M}^{-1}_{\bm{L}}=
\begin{pmatrix}
(1-\bm{w}^T_{\bm{L}} \bm{A}^{-1}_{\bm{L}} \bm{v}_{\bm{L}})^{-1} 
& 
-(1-\bm{w}^T_{\bm{L}} \bm{A}^{-1}_{\bm{L}} \bm{v}_{\bm{L}})^{-1} \bm{w}^T_{\bm{L}} \bm{A}^{-1}_{\bm{L}}
\\
-\bm{A}^{-1}_{\bm{L}} \bm{v}_{\bm{L}}(1-\bm{w}^T_{\bm{L}} \bm{A}^{-1}_{\bm{L}} \bm{v}_{\bm{L}})^{-1}
& 
\bm{A}^{-1}_{\bm{L}}+
\bm{A}^{-1}_{\bm{L}} \bm{v}_{\bm{L}}(1-\bm{w}^T_{\bm{L}} \bm{A}^{-1}_{\bm{L}} \bm{v}_{\bm{L}})^{-1}\bm{w}^T_{\bm{L}} \bm{A}^{-1}_{\bm{L}}
\end{pmatrix}
\eeq
\beq
\det(\bm{M}_{\bm{L}}) 
=
\det(\bm{A}_{\bm{L}}) 
(1-\bm{w}^T_{\bm{L}} \bm{A}^{-1}_{\bm{L}} \bm{v}_{\bm{L}})
.
\eeq
Using this notation, the bias-hardened lensing estimator takes the form
\beq
\hat{\kappa}^\text{BH}_{\bm{L}}
=
\frac{\hat{\kappa}_{\bm{L}} - \bm{w}^T_{\bm{L}} \bm{A}^{-1}_{\bm{L}} \hat{\bm{s}}_{\bm{L}}}
{1-\bm{w}^T_{\bm{L}} \bm{A}^{-1}_{\bm{L}} \bm{v}_{\bm{L}}},
\eeq
where $\hat{\bm{s}}_{\bm{L}} \equiv (\hat{s}_{1,\bm{L}}, \hat{s}_{2,\bm{L}}, \cdots, \hat{s}_{n,\bm{L}})^T$. The variance of the bias-hardened lensing estimator is simply
\beq
\bal
\langle\hat{\kappa}^\text{BH}_{\bm{L}} \hat{\kappa}^\text{BH}_{-\bm{L}}\rangle
=
(1-\bm{w}^T_{\bm{L}} \bm{A}^{-1}_{\bm{L}} \bm{v}_{\bm{L}})^{-2}
\bigg[
\langle\hat{\kappa}_{\bm{L}} \hat{\kappa}_{-\bm{L}} \rangle
&+
\sum_{ij}
\sum_{ab}
(\bm{w}_{\bm{L}})_{i} (\bm{A}^{-1}_{\bm{L}})_{ij} 
(\bm{w}_{\bm{L}})_a (\bm{A}^{-1}_{\bm{L}})_{ab} 
\langle 
\hat{s}_{j,-\bm{L}} 
\hat{s}_{b,\bm{L}}
\rangle
\\
&-
2
\sum_{ij}
(\bm{w}_{\bm{L}})_{i} (\bm{A}^{-1}_{\bm{L}})_{ij}
\text{Re}
\langle\hat{\kappa}_{-\bm{L}}
\hat{s}_{j,\bm{L}} 
\rangle
\bigg]
,
\eal
\eeq
from which we get the noise
\beq
\bal
N^{\kappa^\text{BH}}_{\bm{L}}
=
(1-\bm{w}^T_{\bm{L}} \bm{A}^{-1}_{\bm{L}} \bm{v}_{\bm{L}})^{-2}
\bigg[
N^\kappa_{\bm{L}}
&+
\sum_{ij}
(\bm{w}_{\bm{L}})_{i} (\bm{A}^{-1}_{\bm{L}})_{ij} 
N^{s_j}_{\bm{L}}
\sum_{ab}
(\bm{w}_{\bm{L}})_a (\bm{A}^{-1}_{\bm{L}})_{ab} 
N^{s_b}_{\bm{L}}
\mathcal{R}^{s_b,s_j}_{\bm{L}}
\\
&-
2
N^\kappa_{\bm{L}}
\sum_{ij}
(\bm{w}_{\bm{L}})_{i} (\bm{A}^{-1}_{\bm{L}})_{ij}
N^{s_j}_{\bm{L}}
\mathcal{R}^{\kappa,s_j}_{\bm{L}}
\bigg].
\eal
\eeq
In the past two equations we've assumed that the noises and responses are even functions of $\bm{L}$, which is true so long that the sources are modeled as halos with profiles $u_{\bm{L}}$.
Let's focus on the second term of the RHS of the equation above.
Note that $N^{s_b}_{\bm{L}}
\mathcal{R}^{s_b,s_j}_{\bm{L}}$ is the $b$'th component of the $j$'th column vector of $\bm{A}_{\bm{L}}$. Therefore 
\beq
\sum_b
(\bm{A}^{-1}_{\bm{L}})_{ab} 
N^{s_b}_{\bm{L}}
\mathcal{R}^{s_b,s_j}_{\bm{L}}
=\delta^K_{aj}.
\eeq
From this we find
\beq
\sum_{ab}
(\bm{w}_{\bm{L}})_a (\bm{A}^{-1}_{\bm{L}})_{ab} 
N^{s_b}_{\bm{L}}
\mathcal{R}^{s_b,s_j}_{\bm{L}}
=
\sum_a
(\bm{w}_{\bm{L}})_a 
\sum_b
(\bm{A}^{-1}_{\bm{L}})_{ab} 
N^{s_b}_{\bm{L}}
\mathcal{R}^{s_b,s_j}_{\bm{L}}
=
\sum_a
(\bm{w}_{\bm{L}})_a 
\delta^K_{aj}
=
(\bm{w}_{\bm{L}})_j.
\eeq
Recall that the $j$'th component of $\bm{w}_{\bm{L}}$ is just $N^\kappa_{\bm{L}}\mathcal{R}^{\kappa,s_j}_{\bm{L}}$. Plugging this result back into our expression for the noise gives
\beq
\bal
N^{\kappa^\text{BH}}_{\bm{L}}
&=
(1-\bm{w}^T_{\bm{L}} \bm{A}^{-1}_{\bm{L}} \bm{v}_{\bm{L}})^{-2}
\bigg[
1
-
\sum_{ij}
(\bm{w}_{\bm{L}})_{i} (\bm{A}^{-1}_{\bm{L}})_{ij}
\underbrace{
N^{s_j}_{\bm{L}}
\mathcal{R}^{\kappa,s_j}_{\bm{L}}
}_{=(\bm{v}_{\bm{L}})_j}
\bigg]
N^\kappa_{\bm{L}}
\\
&=
(1-\bm{w}^T_{\bm{L}} \bm{A}^{-1}_{\bm{L}} \bm{v}_{\bm{L}})^{-1}
N^\kappa_{\bm{L}}
\\
&=
\frac{\det(\bm{A}_{\bm{L}})}{\det(\bm{M}_{\bm{L}})}
N^\kappa_{\bm{L}}
.
\eal
\eeq
\section{tSZ-like profile}
\label{app:tsz profile}
\begin{figure}[!h]
\centering
\includegraphics[width=0.3\linewidth]{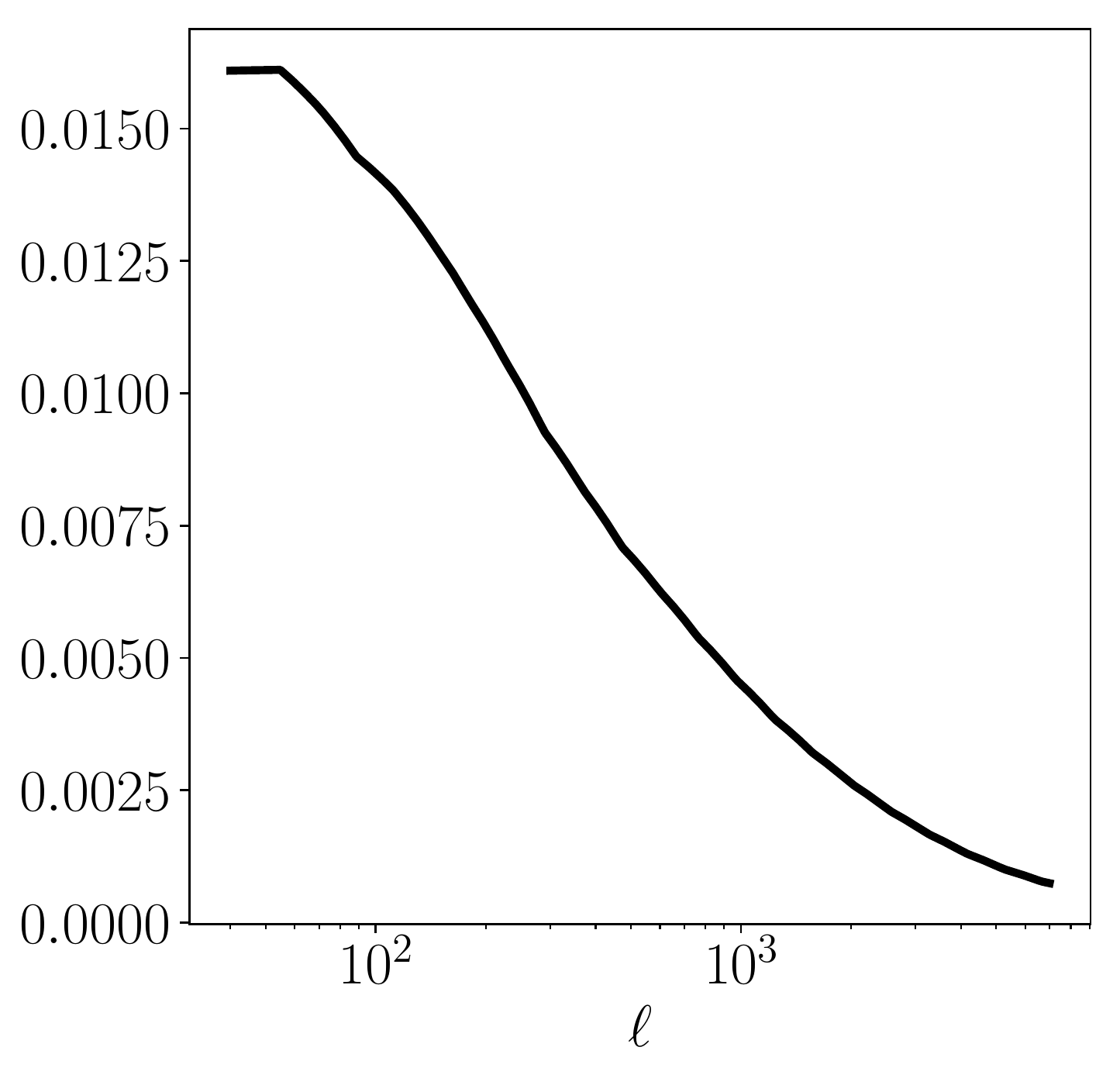}
\caption{ tSZ-like profile $u_{\bm{\ell}}$. For $\ell\lesssim 60$ we approximate the profile as a flat curve.
}
\label{fig:tsz_u}
\end{figure}

If the tSZ foreground map is modeled as $s_{\bm{\ell}} = \sum_i s_i u_{\bm{\ell}}$, then the tSZ power spectrum is proportional to $|u_{\bm{\ell}}|^2$. Recall that the bias-hardened estimator is insensitive to the normalization of the profile. Therefore we take the square root of the tSZ power spectrum (measured from the simulations) as our tSZ-like profile, which is plotted in Fig.~\ref{fig:tsz_u}.
In practice, determining the exact profile from the measured tSZ power spectrum may be challenging.
However, we have checked that the residual biases from tSZ and other foregrounds vary slowly when the assumed profile is changed, such that a precise knowledge of the profile is not needed.
This is encouraging, and suggests that this method will be useful in practice.

\section{Noise price of bias-hardening increases with profile size}
\label{app:noise price for larger profile}

\begin{figure}[!h]
\centering
\includegraphics[width=0.36\linewidth]{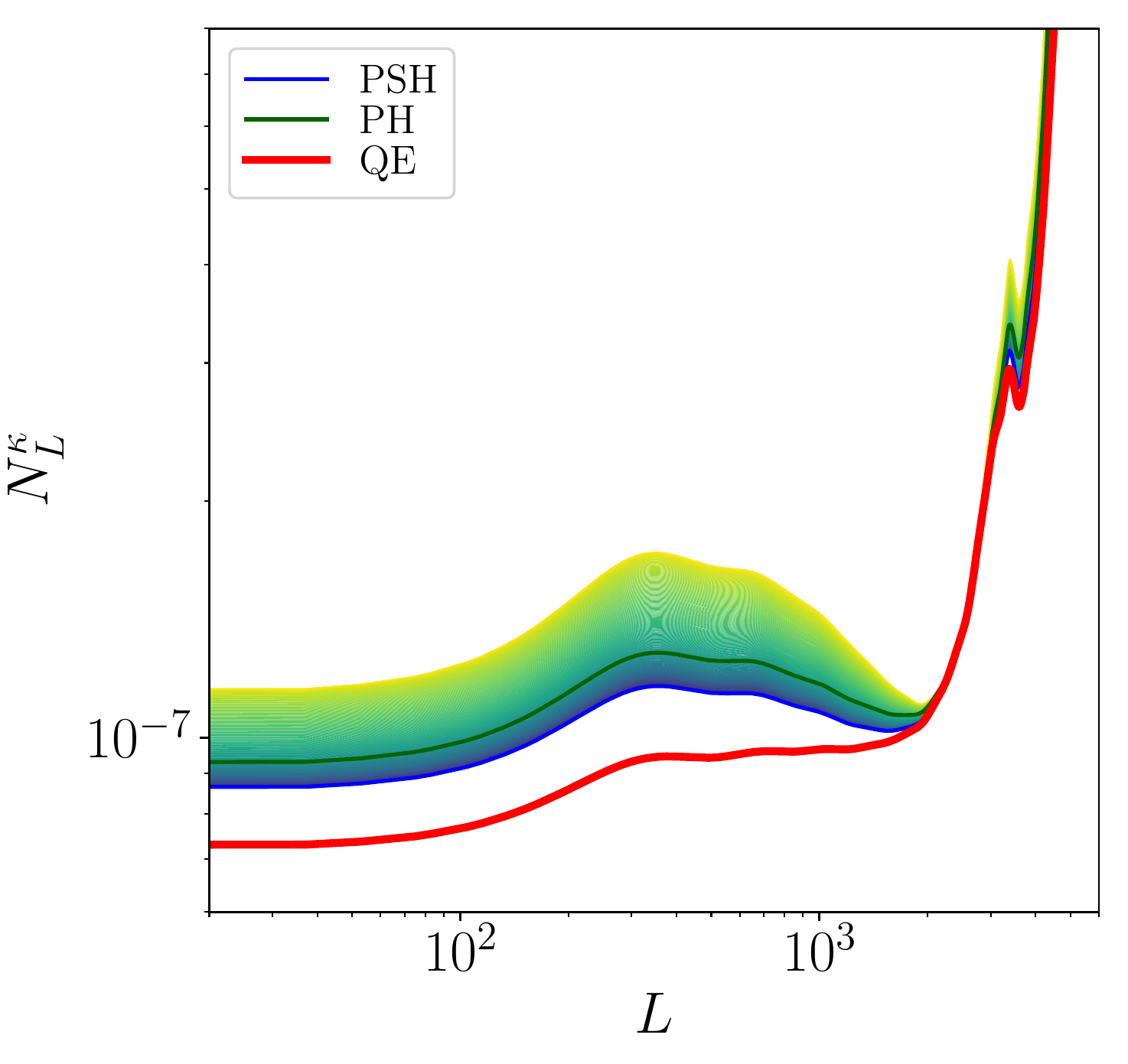}
\caption{ \textbf{Shaded (blue to yellow) region:} Noise when bias hardening against a single Gaussian profile $f^s_{\bm{\ell},\bm{L}-\bm{\ell}} = e^{\sigma^2 L^2 /2}e^{-\sigma^2 \ell^2 /2}e^{-\sigma^2 |\bm{L}-\bm{\ell}|^2 /2}$ for $\sigma\in[0',2.4']$. $\sigma$ is monotonically increasing from blue to yellow. Note that $\sigma=0'$ corresponds to hardening against point sources (PSH). \textbf{Red:} Noise of the standard QE.
}
\label{fig:noise price increases with profile size}
\end{figure}

We found that the noise price for hardening against a tSZ-like profile is larger than hardening against point sources. To gain an intuition for why this is the case, we plot the noise when hardening against a single Gaussian profile $u_{\bm{\ell}} = e^{-\sigma^2 \ell^2 /2}$ for different values of $\sigma$ in Fig.~\ref{fig:noise price increases with profile size}. We find that the noise increases with $\sigma$; that is, the larger the profile the larger the cost in noise. As shown in Fig.~\ref{fig:determinant}, in the regime where the response is small the noise cost goes as $\mathcal{R}^2_L$. The response $\mathcal{R}_L \propto \int d^2 \bm{\ell} f^s_{\bm{\ell},\bm{L}-\bm{\ell}} f^\kappa_{\bm{\ell},\bm{L}-\bm{\ell}}/C^\text{tot}_{\ell}C^\text{tot}_{|\bm{L}-\bm{\ell}|}$ is larger when the linear response to sources $f^s_{\bm{\ell},\bm{L}-\bm{\ell}} = e^{\sigma^2 L^2 /2}e^{-\sigma^2 \ell^2 /2}e^{-\sigma^2 |\bm{L}-\bm{\ell}|^2 /2}$ looks more like $f^\kappa_{\bm{\ell},\bm{L}-\bm{\ell}}$. For point sources $f^s_{\bm{\ell},\bm{L}-\bm{\ell}}$ is flat, however, as the size of the profile gets larger $f^s_{\bm{\ell},\bm{L}-\bm{\ell}}$ looks slightly more similar to the linear response to lensing, making it more difficult for the source-hardened estimator to remove the sources, and results in a higher noise price.

\clearpage

\section{Noise and cross-correlation checks}
\label{app:pipeline_checks}

As a simple pipeline check we verify our calculation of the noise and response when bias-hardening against point sources (PSH). These checks are shown in the plots below.

\begin{figure}[!h]
\centering
\includegraphics[width=.3\linewidth]{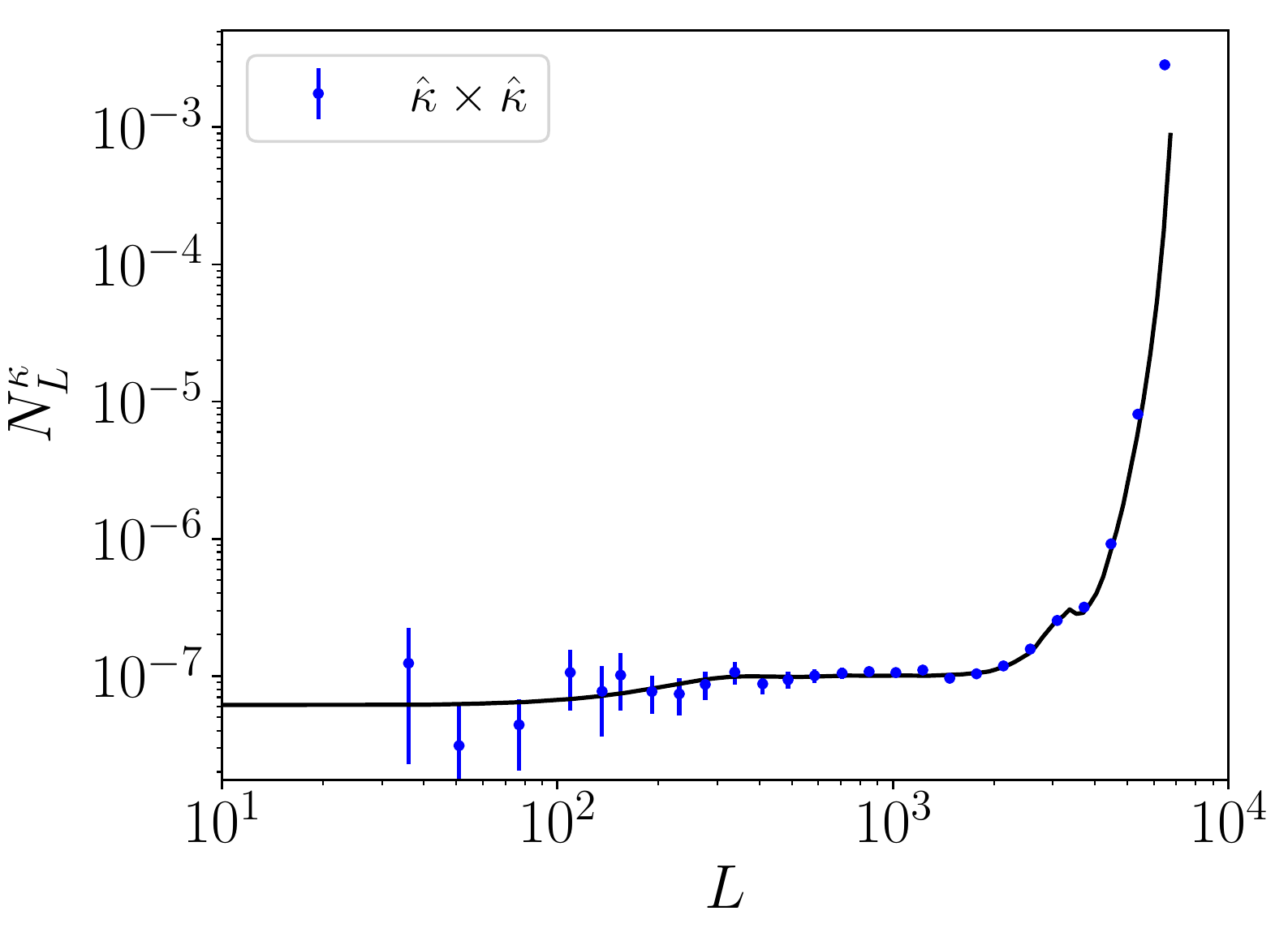}
\includegraphics[width=.3\linewidth]{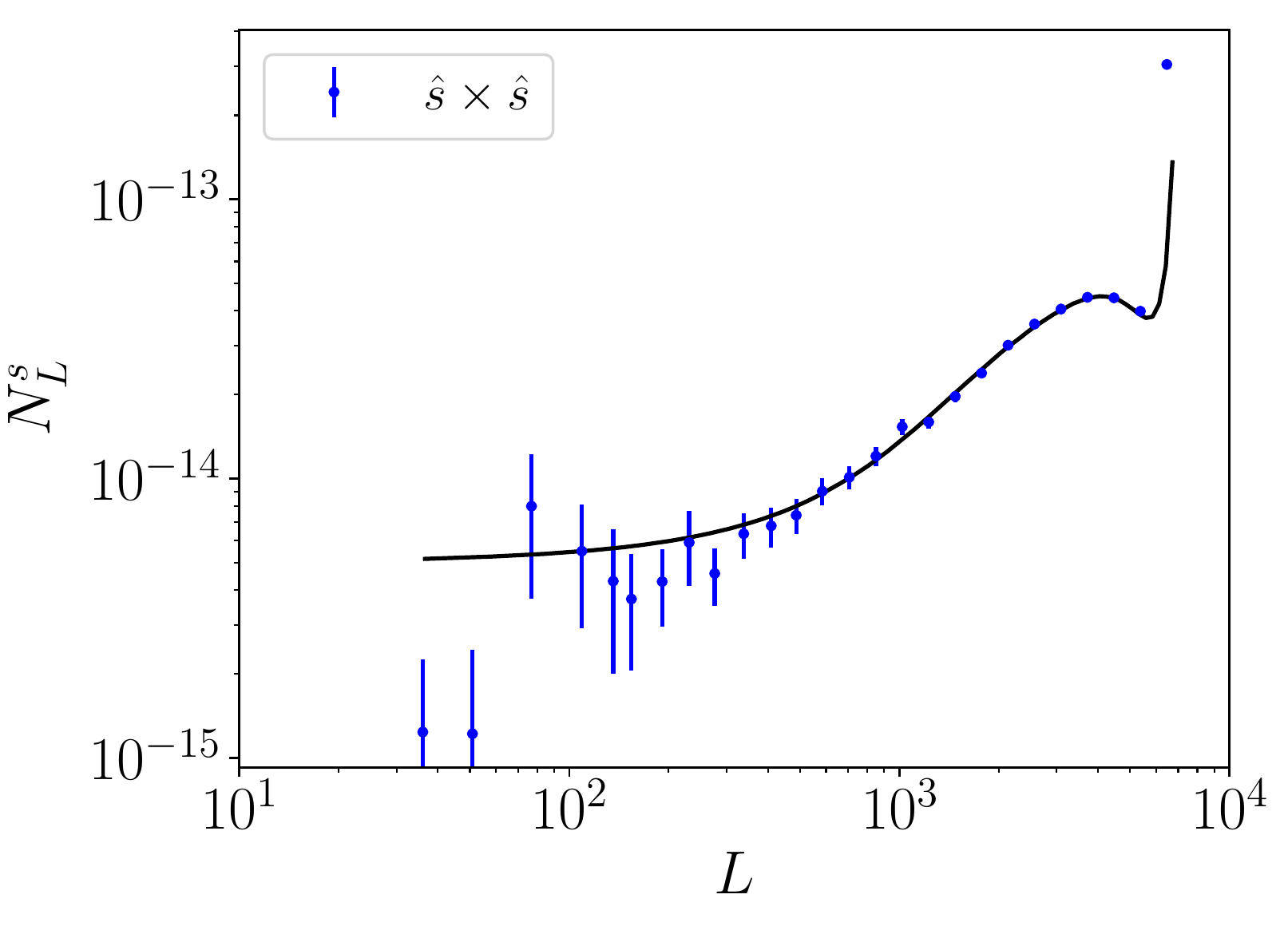}
\includegraphics[width=.3\linewidth]{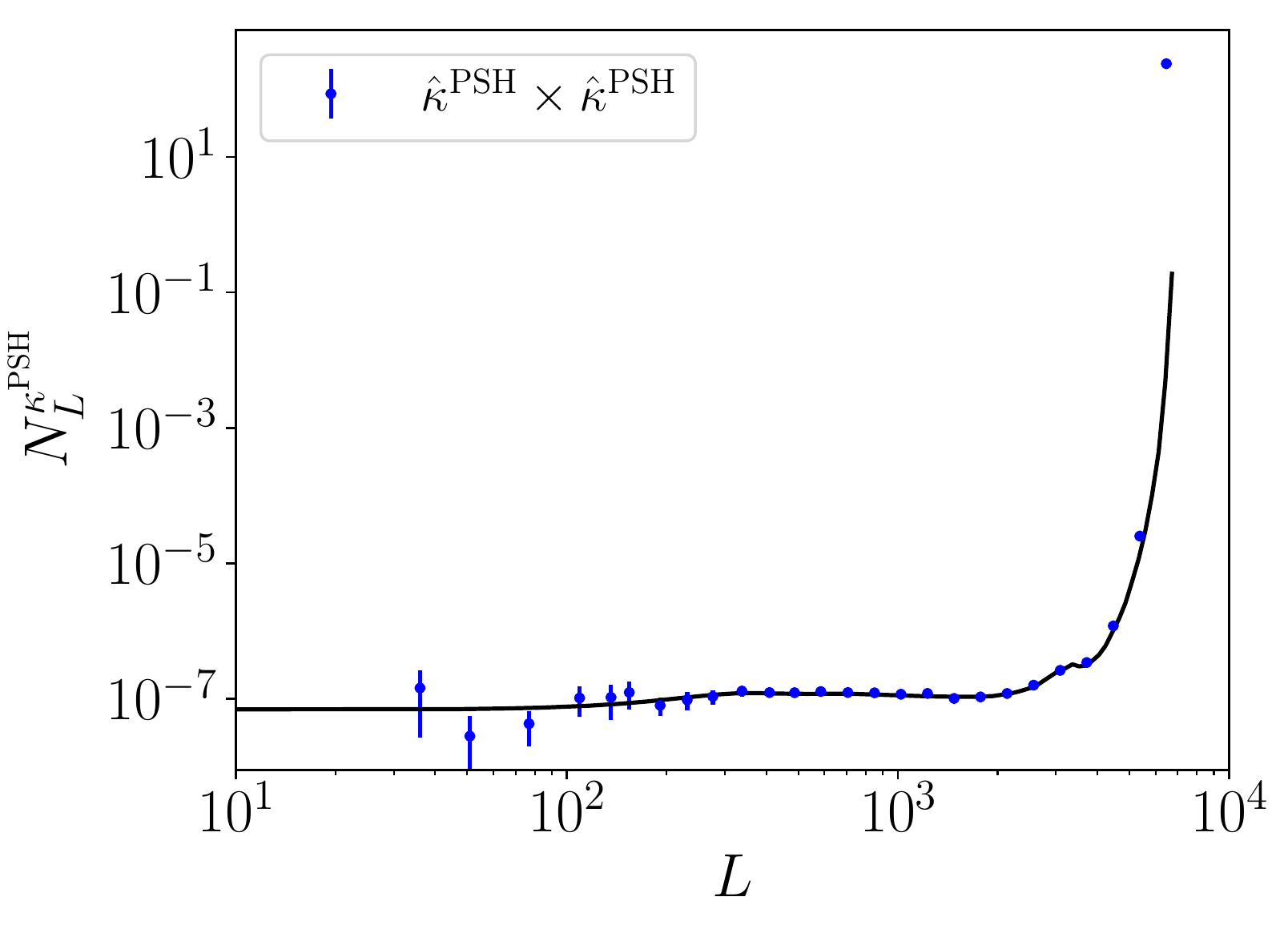}
\caption{Verification of the noises of each estimator. The auto-correlations (blue) of the estimators $\hat{\kappa},\hat{s},\text{ and }\hat{\kappa}^\text{PSH}$ when run on a Gaussian random field with power spectrum $C^\text{tot}_L$. In black are the theory curves.
}
\end{figure}

\begin{figure}[!h]
\centering
\includegraphics[width=.3\linewidth]{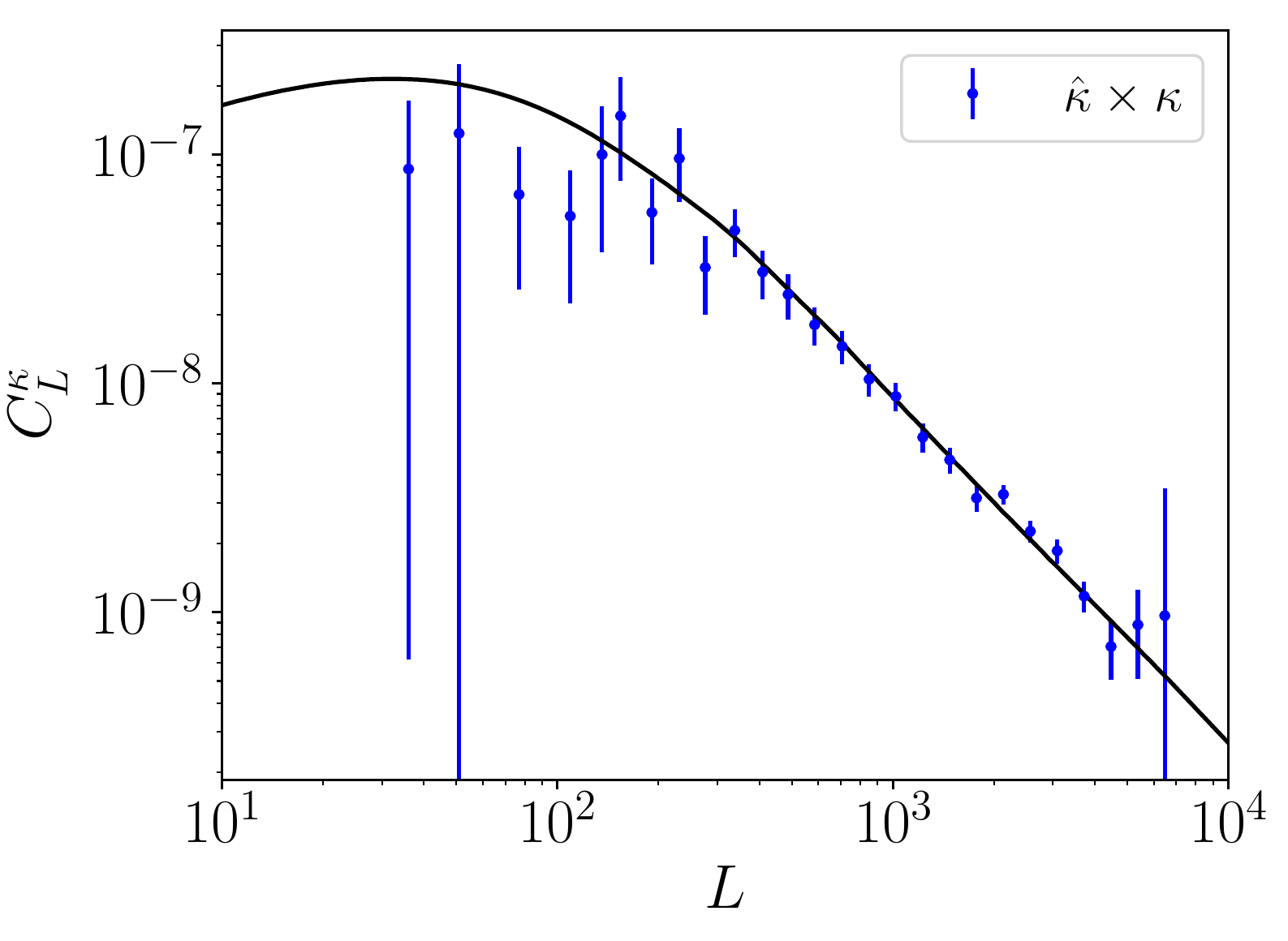}
\includegraphics[width=.3\linewidth]{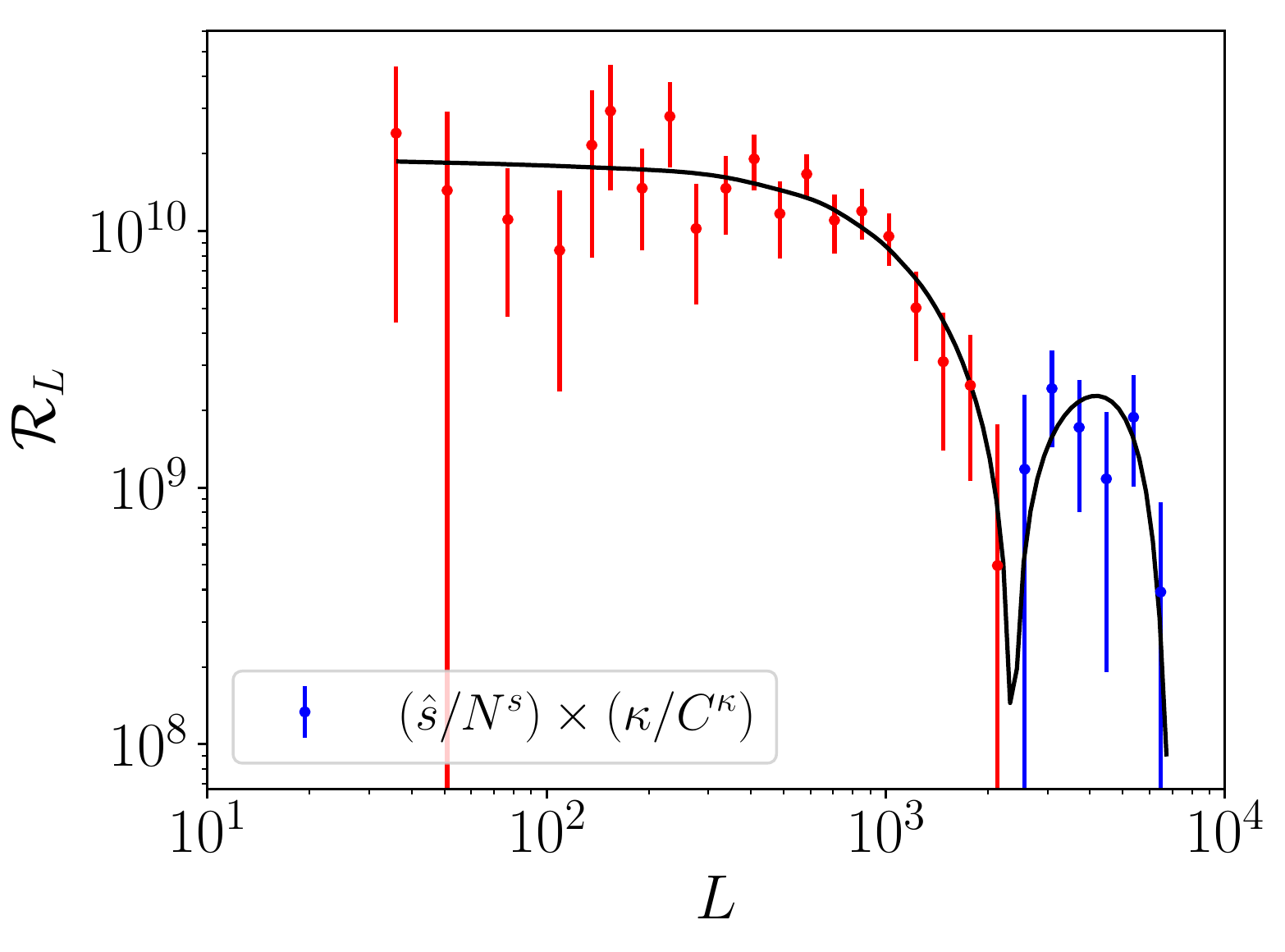}
\includegraphics[width=.3\linewidth]{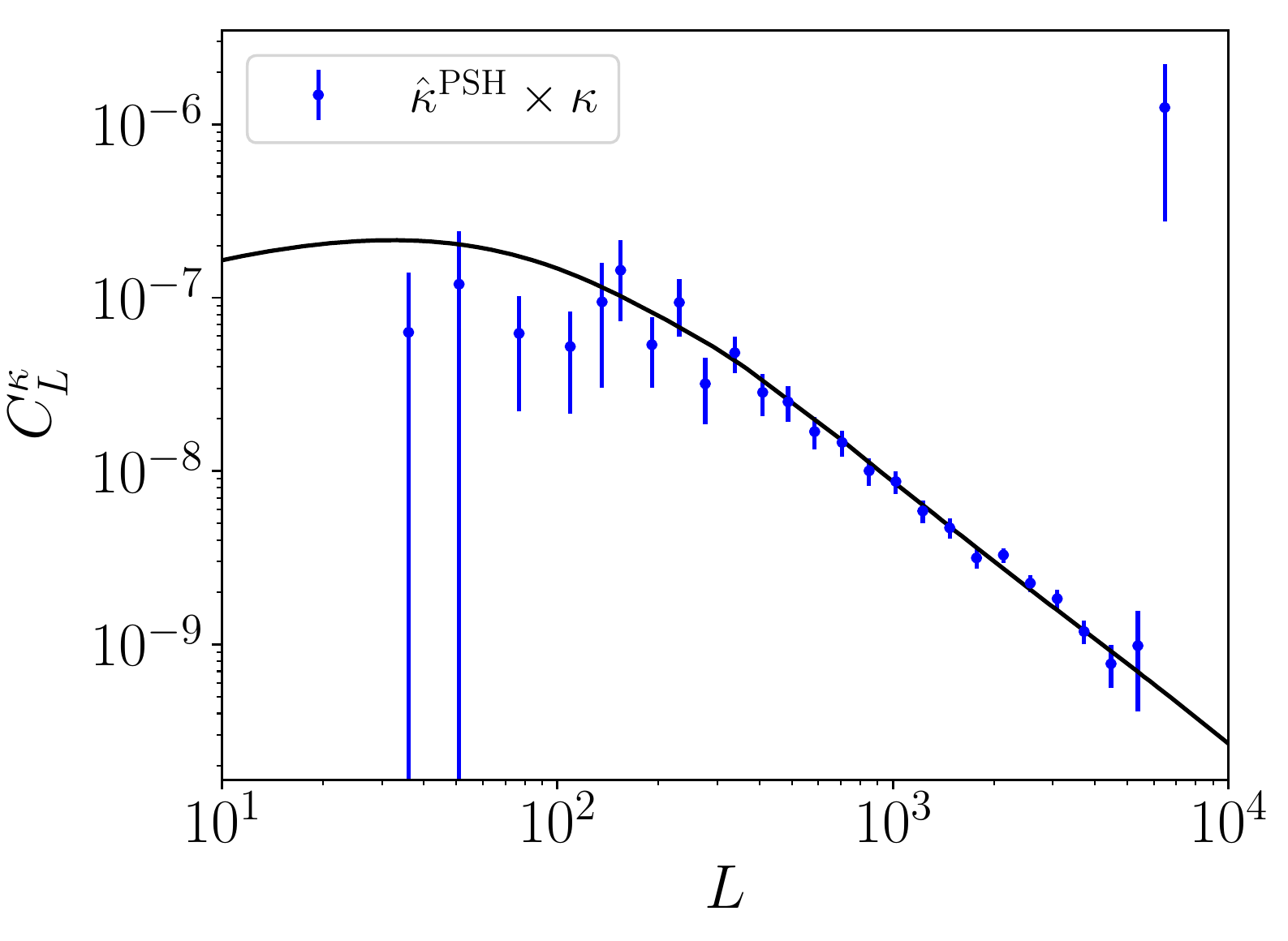}
\caption{Cross-correlating the true convergence $\kappa$ with $\hat{\kappa}$, $\hat{s}$, and $\hat{\kappa}^\text{BH}$ when run on a lensed CMB map. We see that both the standard QE $\hat{\kappa}$ and the point source hardened estimator $\hat{\kappa}^\text{PSH}$ recover the lensing signal $C^\kappa_L$ when correlated with the true convergence. The cross correlation of the point source estimator $\hat{s}$ with the true lensing convergence gives $N^s_L C^\kappa_L \mathcal{R}_L$, which is verified in the middle plot. The blue points are positive, whereas the red points are negative. The response has a zero crossing at $L\sim2500$.}
\end{figure}

\begin{figure}[!h]
\centering
\includegraphics[width=.3\linewidth]{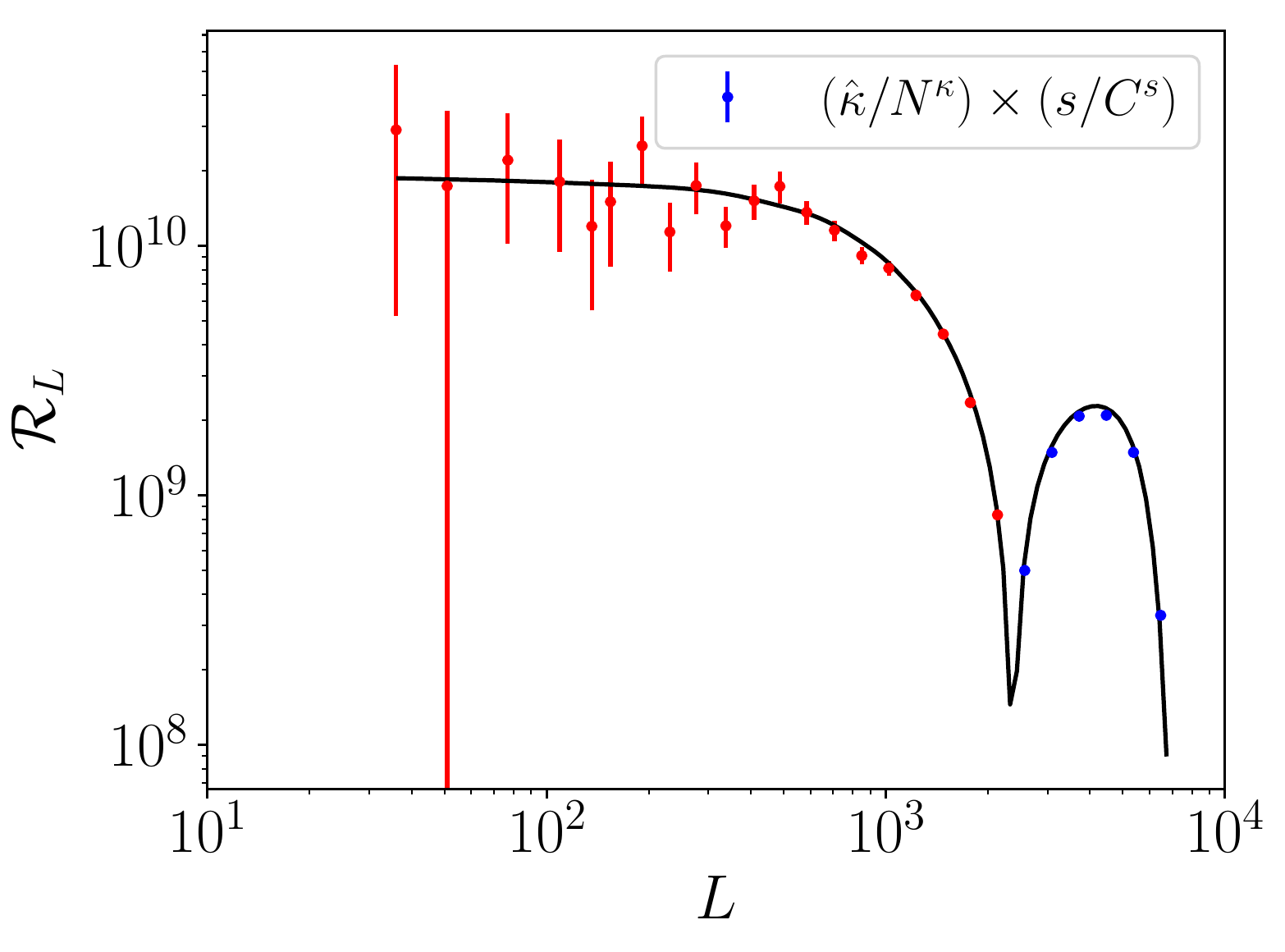}
\includegraphics[width=.3\linewidth]{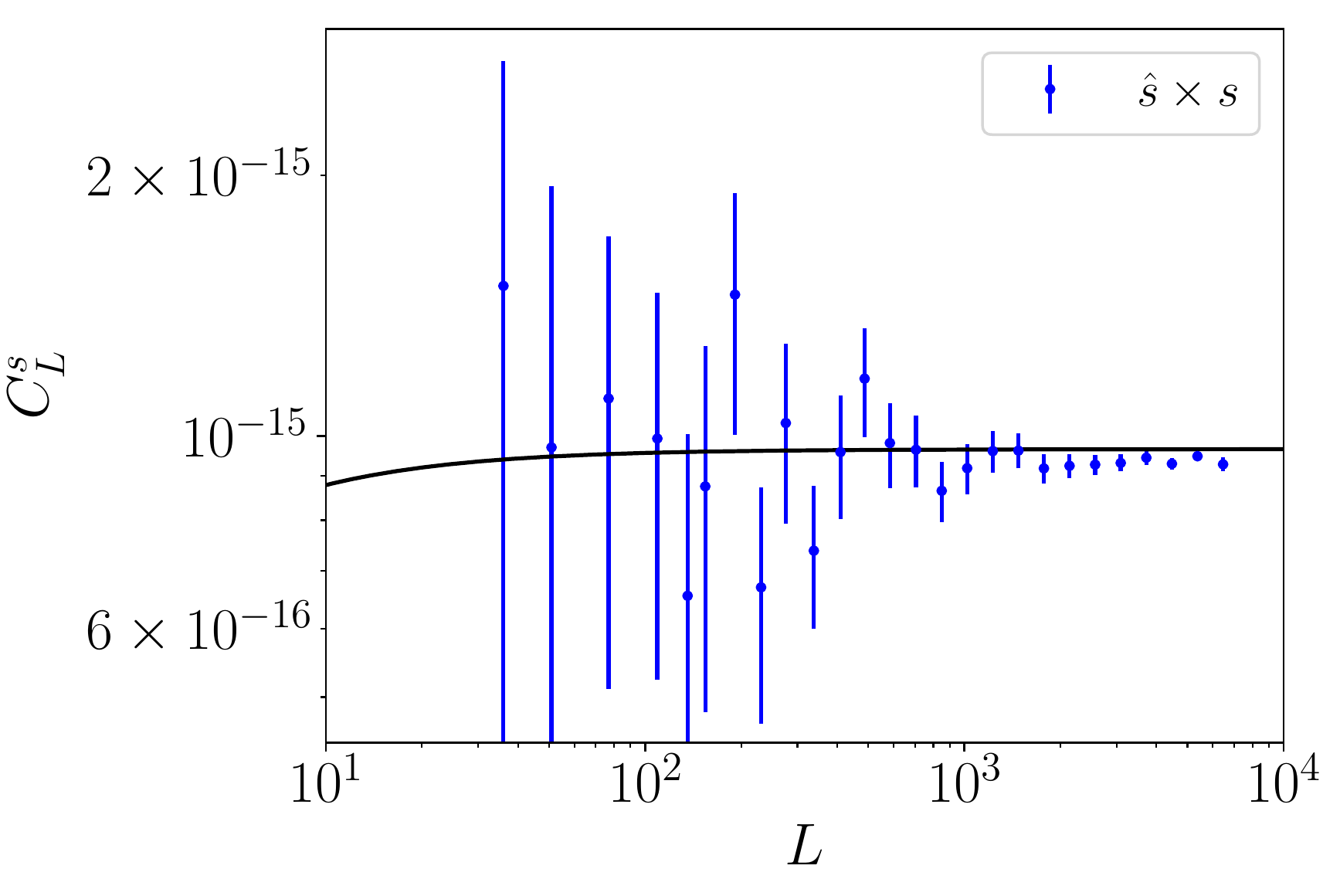}
\includegraphics[width=.3\linewidth]{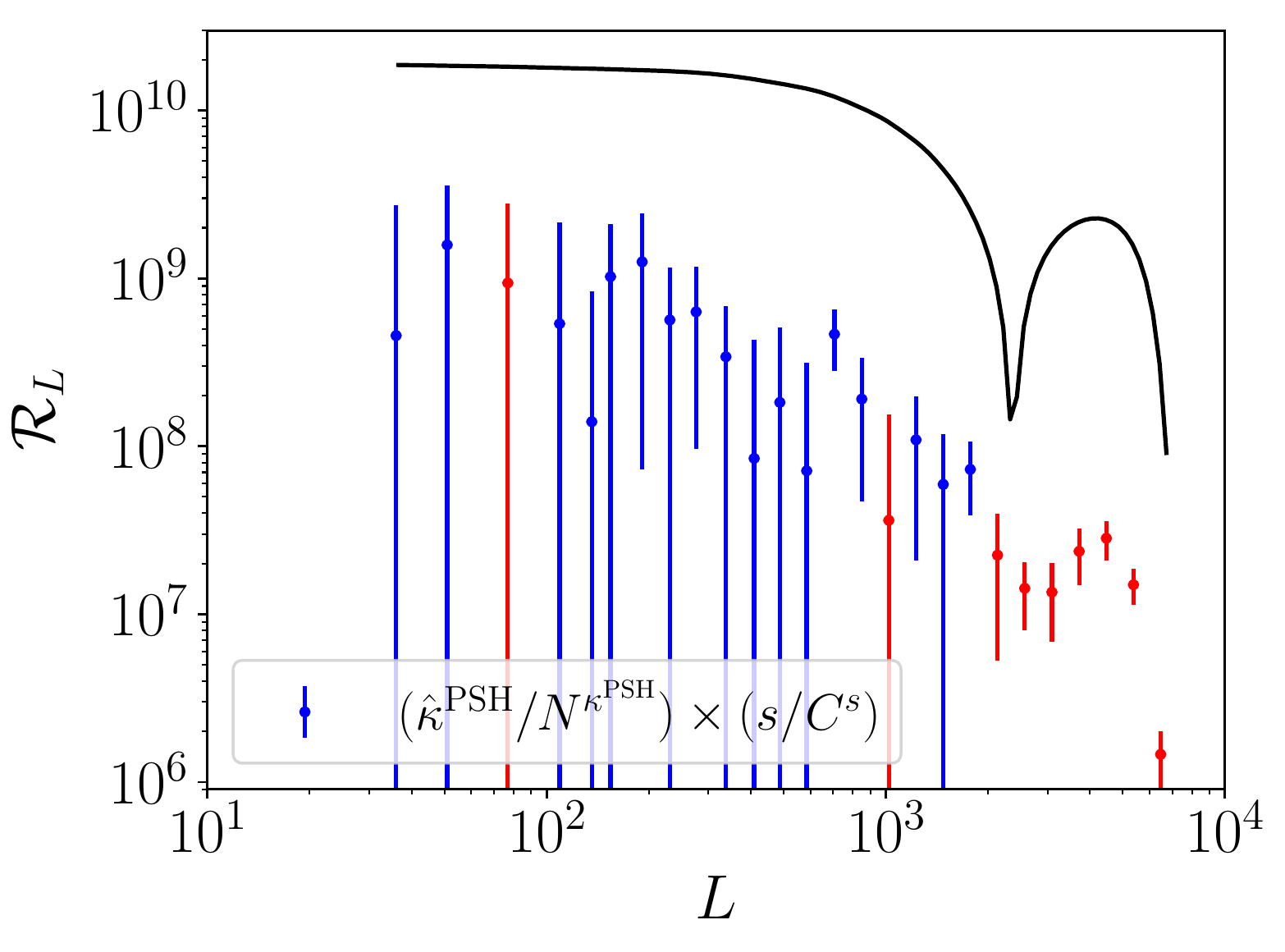}
\caption{Cross-correlating the true source map $s$ with $\hat{\kappa}$, $\hat{s}$, and $\hat{\kappa}^\text{PSH}$ when run on the true point source map $s$. The cross-correlation of $\hat{\kappa}$ with $s$ gives $N^\kappa_L C^s_L \mathcal{R}_L$, which is verified in the plot on the left. $\hat{s}$ recovers the source signal $C^s_L$ when cross correlated with the true source map. Since the point-source hardened estimator is designed to have zero response to point sources, the cross-correlation of $\hat{\kappa}^\text{PSH}$ and $s$ should be zero, which is consistent with the result found in the rightmost plot.}
\end{figure}

\clearpage
\bibliographystyle{prsty.bst}
\bibliography{main}

\begin{thebibliography}{10}

\bibitem{2006PhR...429....1L}
A. {Lewis} and A. {Challinor}, \physrep {\bf 429},  1  (2006).

\bibitem{2010GReGr..42.2197H}
D. {Hanson}, A. {Challinor}, and A. {Lewis}, General Relativity and Gravitation
  {\bf 42},  2197  (2010).

\bibitem{2016JLTP..184..772H}
S.~W. {Henderson} {\it et~al.}, Journal of Low Temperature Physics {\bf 184},
  772  (2016).

\bibitem{2014SPIE.9153E..1PB}
B.~A. {Benson} {\it et~al.},  {\bf 9153},  91531P  (2014).

\bibitem{2019JCAP...02..056A}
P. {Ade} {\it et~al.}, \jcap {\bf 2019},  056  (2019).

\bibitem{2019arXiv190704473A}
K. {Abazajian} {\it et~al.}, arXiv e-prints  arXiv:1907.04473  (2019).

\bibitem{sehgal2019cmbhd}
N. Sehgal {\it et~al.}, CMB-HD: An Ultra-Deep, High-Resolution Millimeter-Wave
  Survey Over Half the Sky, 2019.

\bibitem{sehgal2020cmbhd}
N. Sehgal {\it et~al.}, CMB-HD: Astro2020 RFI Response, 2020.

\bibitem{2014ApJ...786...13V}
A. {van Engelen} {\it et~al.}, \apj {\bf 786},  13  (2014).

\bibitem{2014JCAP...03..024O}
S.~J. {Osborne}, D. {Hanson}, and O. {Dor{\'e}}, \jcap {\bf 2014},  024
  (2014).

\bibitem{2018PhRvD..98b3534M}
M.~S. {Madhavacheril} and J.~C. {Hill}, \prd {\bf 98},  023534  (2018).

\bibitem{2018PhRvD..97b3512F}
S. {Ferraro} and J.~C. {Hill}, \prd {\bf 97},  023512  (2018).

\bibitem{2019PhRvL.122r1301S}
E. {Schaan} and S. {Ferraro}, \prl {\bf 122},  181301  (2019).

\bibitem{2019PhRvD..99b3508B}
E.~J. {Baxter} {\it et~al.}, \prd {\bf 99},  023508  (2019).

\bibitem{2020arXiv200401139D}
O. {Darwish} {\it et~al.}, arXiv e-prints  arXiv:2004.01139  (2020).

\bibitem{2013MNRAS.431..609N}
T. {Namikawa}, D. {Hanson}, and R. {Takahashi}, \mnras {\bf 431},  609  (2013).

\bibitem{plancklensing2013}
P.~A.~R. Ade {\it et~al.}, Astronomy \& Astrophysics {\bf 571},  A17  (2014).

\bibitem{2002ApJ...574..566H}
W. {Hu} and T. {Okamoto}, \apj {\bf 574},  566  (2002).

\bibitem{Fabbian_2019}
G. Fabbian, A. Lewis, and D. Beck, Journal of Cosmology and Astroparticle
  Physics {\bf 2019},  057–057  (2019).

\bibitem{2010ApJ...709..920S}
N. {Sehgal} {\it et~al.}, \apj {\bf 709},  920  (2010).

\bibitem{2018JCAP...07..046F}
S. {Foreman}, P.~D. {Meerburg}, A. {van Engelen}, and J. {Meyers}, \jcap {\bf
  2018},  046  (2018).

\bibitem{2013JCAP...07..025D}
J. {Dunkley} {\it et~al.}, \jcap {\bf 2013},  025  (2013).

\bibitem{2009arXiv0912.0201L}
{LSST Science Collaboration} {\it et~al.}, arXiv e-prints  arXiv:0912.0201
  (2009).

\end{thebibliography}

\end{document}